\documentclass[preprint,12pt]{imsart}
\usepackage{amssymb,amsmath,bm}
\usepackage{url}
\usepackage{dcolumn} 
\usepackage{bm} 
\usepackage{graphicx}

\usepackage{natbib}
\def\cite{\citet}

\DeclareMathOperator*{\argmax}{arg\,max}
\DeclareMathOperator*{\argmin}{arg\,min}

\usepackage{geometry}
\begin{document}

\begin{frontmatter}
\title{Selective Inference for Testing Trees and Edges in Phylogenetics}
\runtitle{Selective Inference for Phylogenetics}
\author{\fnms{Hidetoshi} \snm{Shimodaira$^{1,3}$}}
\and
\author{\fnms{Yoshikazu} \snm{Terada$^{2,3}$}} 

\address{\texttt{shimo@i.kyoto-u.ac.jp} and \texttt{terada@sigmath.es.osaka-u.ac.jp}\\
$^{1}$Graduate School of Informatics, Kyoto University, Yoshida Honmachi, Sakyo-ku, Kyoto, 606-8501, Japan\\
$^{2}$Graduate School of Engineering Science, Osaka University, 1-3 Machikaneyama-cho, Toyonaka, Osaka 560-8531, Japan\\
$^{3}$Mathematical Statistics Team, RIKEN Center for Advanced Intelligence Project, 1-4-1 Nihonbashi, Chuo-ku, Tokyo 103-0027, Japan}
 
\runauthor{H.~SHIMODAIRA AND Y.~TERADA}

\begin{abstract}


Selective inference is considered for testing trees and edges in phylogenetic tree selection from molecular sequences.
This improves the previously proposed approximately unbiased test by adjusting the selection bias when testing many trees and edges at the same time.
The newly proposed selective inference $p$-value is useful for testing selected edges to claim that they are significantly supported if $p>1-\alpha$, whereas the non-selective $p$-value is still useful for testing candidate trees to claim that they are rejected if $p<\alpha$.
The selective $p$-value controls the type-I error conditioned on the selection event, whereas the non-selective $p$-value controls it unconditionally.
The selective and non-selective approximately unbiased $p$-values are computed from two geometric quantities called signed distance and mean curvature of the region representing tree or edge of interest in the space of probability distributions. These two geometric quantities are estimated by fitting a model of scaling-law to the non-parametric multiscale bootstrap probabilities.
Our general method is applicable to a wider class of problems; phylogenetic tree selection is an example of model selection, and it is interpreted as the variable selection of multiple regression, where each edge corresponds to each predictor.
Our method is illustrated in a previously controversial phylogenetic analysis of human, rabbit and mouse.



\end{abstract}


\end{frontmatter}





\section{Introduction} \label{sec:intro}


A phylogenetic tree is a diagram showing evolutionary relationships among species, and a tree topology is a graph obtained from the phylogentic tree by ignoring the branch lengths.
The primary objective of any phylogenetic analysis is to approximate a topology that reflects the evolution history of the group of organisms under study.
Branches of the tree are also referred to as edges in the tree topology.
Given a rooted tree topology, or a unrooted tree topology with an outgroup,  each edge splits the tree so that it defines the clade consisting of all the descendant species.
Therefore, edges in a tree topology represent clades of species.
Because the phylogenetic tree is commonly inferred from molecular sequences, it is crucial to assess the statistical confidence of the inference.
In phylogenetics, it is a common practice to compute confidence levels for tree topologies and edges.
For example, the bootstrap probability~\citep{Felsenstein:1985:CLP} is the most commonly used confidence measure, and other methods such as
the Shimodaira-Hasegawa test~\citep{shimodaira1999multiple}
and the multiscale bootstrap method~\citep{shimodaira2002approximately} are also often used.
However, these conventional methods are limited in how well they address the issue of multiplicity when there are many alternative topologies and edges.  
Herein, we discuss a new approach, selective inference (SI), that is designed to address the issue of multiplicity.

For illustrating the idea of selective inference, we first look at a simple example of 1-dimensional normal random variable $Z$ with unknown mean $\theta\in\mathbb{R}$ and variance 1:
\begin{equation} \label{eq:z-normal}
    Z \sim N(\theta,1).
\end{equation}
Observing $Z=z$, we would like to test the null hypothesis $H_0: \theta\le 0$ against the alternative hypothesis $H_1: \theta>0$. 
We denote the cumulative distribution function of $N(0,1)$ as $\Phi(x)$ and define the upper tail probability as $\bar\Phi(x) = 1 - \Phi(x) = \Phi(-x)$.
Then, the ordinary (i.e., non-selective) inference leads to the $p$-value of the one-tailed $z$-test as
\begin{equation} \label{eq:z-au}
    p(z) := P(Z > z \mid \theta=0) = \bar\Phi(z).
\end{equation}
What happens when we test many hypotheses at the same time?
Consider random variables $Z_i \sim N(\theta_i, 1)$, $i=1,\ldots, K_\text{all}$, not necessarily independent, with null hypotheses $\theta_i\le 0$, where $K_\text{true}$ hypotheses are actually true.
To control the number of falsely rejecting the $K_\text{true}$ hypotheses, there are several multiplicity adjusted approaches such as the family-wise error rate (FWER) and the false discovery rate (FDR).
Instead of testing all the $K_\text{all}$ hypotheses, selective inference (SI) allows for $K_\text{select}$ hypotheses with $z_i > c_i$ for constants $c_i$ specified in advance.
This kind of selection is very common in practice (e.g., publication bias), and it is called as the \emph{file drawer problem} by \cite{rosenthal1979file}.
Instead of controlling the multiplicity of testing, SI alleviates it by reducing the number of tests.
The mathematical formulation of SI is easier than FWER and FDR in the sense that hypotheses can be considered separately instead of simultaneously.
Therefore, we simply write $z>c$ by dropping the index $i$ for one of the hypotheses.
In selective inference, the selection bias is adjusted by considering the conditional probability given the selection event,
which leads to the following $p$-value~\citep{fithian2014optimal,tian2018selective}
\begin{equation} \label{eq:z-si}
     p(z, c) := P(Z > z \mid Z >c, \theta=0) = \bar\Phi(z) / \bar\Phi(c),
\end{equation}
where $p(z)$ of eq.~(\ref{eq:z-au}) is divided by the selection probability $P(Z>c \mid \theta=0) = \bar\Phi(c)$.
In the case of $c=0$, this corresponds to the two-tailed $z$-test, because the selection probability is $\bar\Phi(0)=0.5$ and  $p(z,c)= 2p(z)$.
For significance level $\alpha$ (we use $\alpha=0.05$ unless otherwise stated), it properly controls the type-I error conditioned on the selection event as
$P(p(Z,c) < \alpha \mid Z>c, \theta=0) = \alpha$, while the non-selective $p$-value violates the type-I error as
$P(p(Z) < \alpha \mid Z>c, \theta=0) = \alpha/\bar\Phi(c) >\alpha$. 
The selection bias can be very large when $\bar\Phi(c) \ll 1$ (i.e. $c\gg0$), or $K_\text{select} \ll K_\text{all}$.



\begin{figure}[h!]
\begin{center}
\includegraphics[width=5cm]{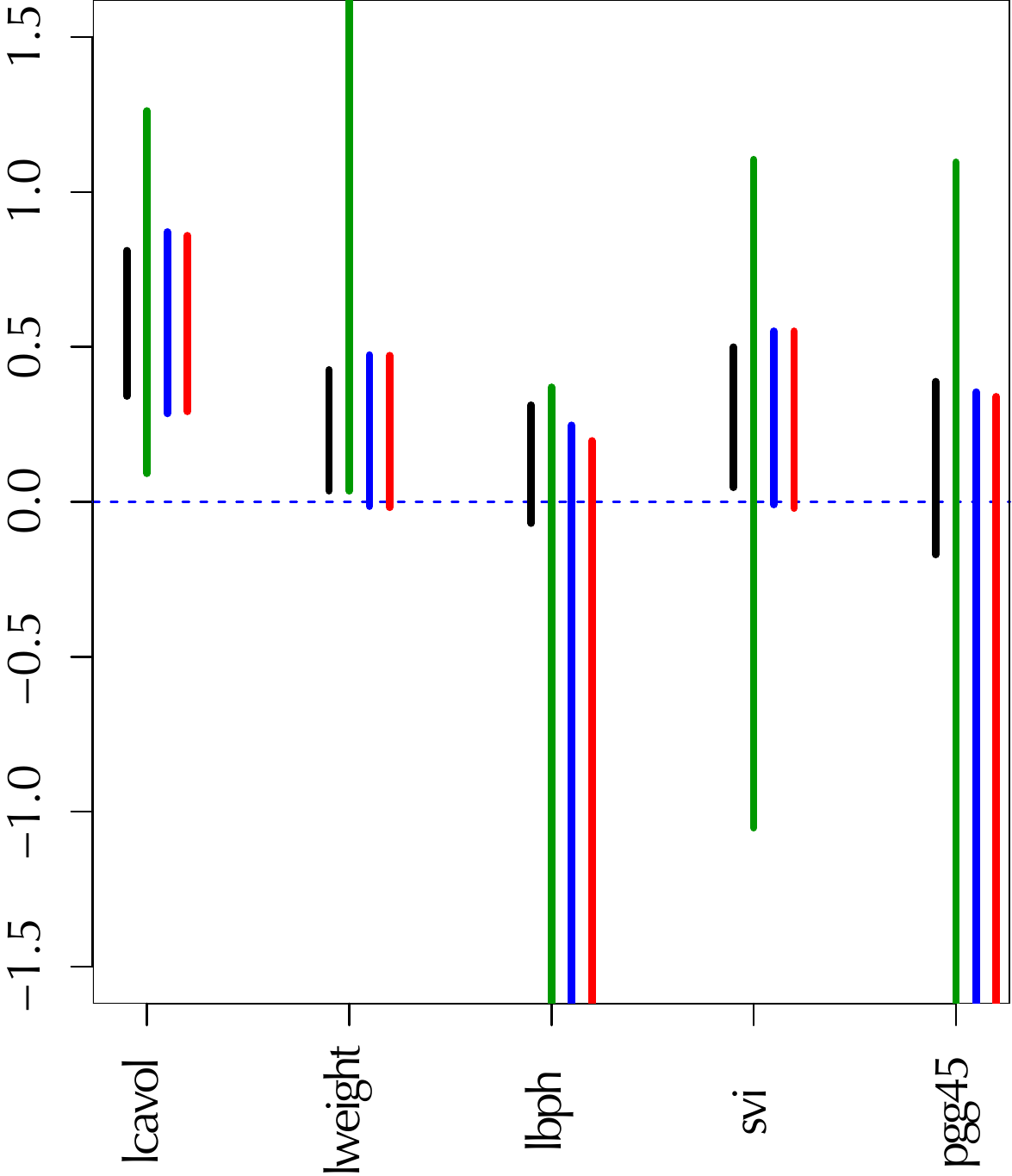}
\end{center}
\caption{Confidence intervals of regression coefficients for selected variables by lasso;  see Section~\ref{sec:si_lasso} for details.
All intervals are computed for confidence level $1-\alpha$ at $\alpha=0.01$.
(Black)~the ordinary confidence interval $[L_j^\text{ordinary}, U_j^\text{ordinary}]$.
(Green)~the selective confidence interval $[L_j^\text{model}, U_j^\text{model}]$ under the selected model.
(Blue)~the selective confidence interval $[L_j^\text{variable}, U_j^\text{variable}]$ under the selection event that variable $j$ is selected.
(Red)~the multiscale bootstrap version of selective confidence interval
$[\hat L_j^\text{variable}, \hat U_j^\text{variable}]$ under the selection event that variable $j$ is selected.
}\label{fig:lasso}
\end{figure}

Selective inference has been mostly developed for inferences after model selection~\citep{TaylorTibshirani15,RyanTibshiraniEtAl16}, particularly variable selection in regression settings such as lasso~\citep{tibshirani1996regression}.
Recently, \cite{terada2017selective} developed a general method for selective inference by adjusting
the selection bias in the approximately unbiased (AU) $p$-value computed by the multiscale bootstrap method~\citep{shimodaira2002approximately,shimodaira2004approximately,shimodaira2008testing}.
This new method can be used to compute, for example, confidence intervals of regression coefficients in lasso (figure~\ref{fig:lasso}).
In this paper, we apply this method to phylogenetic inference for computing proper confidence levels of tree topologies (dendrograms) and edges (clades or clusters) of species.
As far as we know, this is the first attempt to consider selective inference in phylogenetics.
Our selective inference method is implemented in software \emph{scaleboot}~\citep{shimo2019scaleboot} working jointly with \emph{CONSEL}~\citep{shimodaira2001consel} for phylogenetics,
and it is also implemented in a new version of \emph{pvclust}~\citep{suzuki2006pvclust} for hierarchical clustering,
where only edges appeared in the observed tree are ``selected'' for computing $p$-values.
Although our argument is based on the rigorous theory of mathematical statistics in \cite{terada2017selective},
a self-contained illustration is presented in this paper for 
the theory as well as the algorithm of selective inference.

Phylogenetic tree selection is an example of model selection.
Since each tree can be specified as a combination of edges,
tree selection can be interpreted as the variable selection of multiple regression,
where edges correspond to the predictors of regression~\citep{shimodaira2001multiple,shimodaira2005assessing}.
Because all candidate trees have the same number of model parameters, the maximum likelihood (ML) tree is obtained by comparing log-likelihood values of trees~\citep{bib:Fels:81:ETD}.
In order to adjust the model complexity by the number of parameters in general model selection,
we compare Akaike Information Criterion (AIC) values of candidate models~\citep{akaike1974new}. AIC is used in phylogenetics for selecting the substitution model~\citep{posada2004model}.
There are several modifications of AIC that allow for model selection.  These include the precise estimation of
the complexity term known as Takeuchi Information Criterion \citep{burnham2002model,konishi2008information}, and adaptations for incomplete data \citep{shimodaira2018information}
and covariate-shift data \citep{shimodaira2000improving}.
AIC and all these modifications are derived for estimating the expected Kullback-Leibler divergence between the unknown true distribution and the estimated probability distribution on the premise that the model is misspecified.
When using regression model for prediction purpose, it may be sufficient to find only the best model which minimizes the AIC value.
Considering random variations of dataset, however, it is obvious in phylogenetics that the ML tree does not necessarily represent the true history of evolution.
Therefore, \cite{bib:Kish:Hase:89:EML} proposed a statistical test whether two log-likelihood values differ significantly (also known as \emph{Kishino-Hasegawa} test).
The log-likelihood difference is often not significant, because its variance can be very large for non-nested models when the divergence between two probability distributions is large; see eq.~(\ref{eq:var-lik}) in Section~\ref{sec:rell}.
The same idea of model selection test whether two AIC values differ significantly has been proposed independently in statistics
\citep{linhart1988testaic} and econometrics \citep{vuong1989likelihood}.
Another method of model selection test \citep{efron1984comparing} allows for the comparison of two regression models with an adjusted bootstrap confidence interval corresponding to the AU $p$-value.
For testing which model is better than the other, the null hypothesis in the model selection test is that the two models are equally good in terms of the expected value of AIC on the premise that both models are misspecified.
Note that the null hypothesis is whether the model is correctly specified or not in the traditional hypothesis testing methods including the likelihood ratio test for nested models and the modified likelihood ratio test for non-nested models \citep{cox1962further}.
The model selection test is very different from these traditional settings.
For comparing AIC values of more than two models, a multiple comparisons method is introduced to the model selection test
\citep{shimodaira1998application,shimodaira1999multiple}, which computes the confidence set of models.
But the multiple comparisons method is conservative by nature, leading to more false negatives than expected, because it considers the worst scenario, called the least favorable configuration.
On the other hand, the model selection test (designed for two models) and bootstrap probability \citep{Felsenstein:1985:CLP} lead to more false positives than expected when comparing more than two models \citep{shimodaira1999multiple,shimodaira2002approximately}.
The AU $p$-value mentioned earlier has been developed for solving this problem, and we are going to upgrade it for selective inference.

\section{Phylogenetic Inference} \label{sec:phylo}

For illustrating phylogenetic inference methods, we analyze a dataset consisting of mitochondrial protein sequences of six mammalian species with $n=3414$ amino acids ($n$ is treated as sample size).
The taxa are labelled as 1$=${\em Homo sapiens} (human), 2$=${\em Phoca vitulina} (seal), 3$=${\em Bos taurus} (cow), 4$=${\em Oryctolagus cuniculus} (rabbit), 5$=${\em Mus musculus} (mouse), and 6$=${\em Didelphis virginiana} (opossum).
The dataset will be denoted as $\mathcal{X}_n = (\bm{x}_1,\ldots,\bm{x}_n)$.
The software package PAML \citep{bib:Yang:97:PPP} was used to calculate the site-wise log-likelihoods for trees.  The mtREV model \citep{bib:Adac:Hase:96:MAA} was used for amino acid substitutions, and the site-heterogeneity was modeled by the discrete-gamma distribution \citep{bib:Yang:96:ASR}.
The dataset and evolutionary model are similar to previous publications \citep{shimodaira1999multiple,shimodaira2001multiple,shimodaira2002approximately}, thus allowing our proposed method to be easily compared with conventional methods.

\begin{figure}[h!]
\begin{center}
\includegraphics[width=10cm]{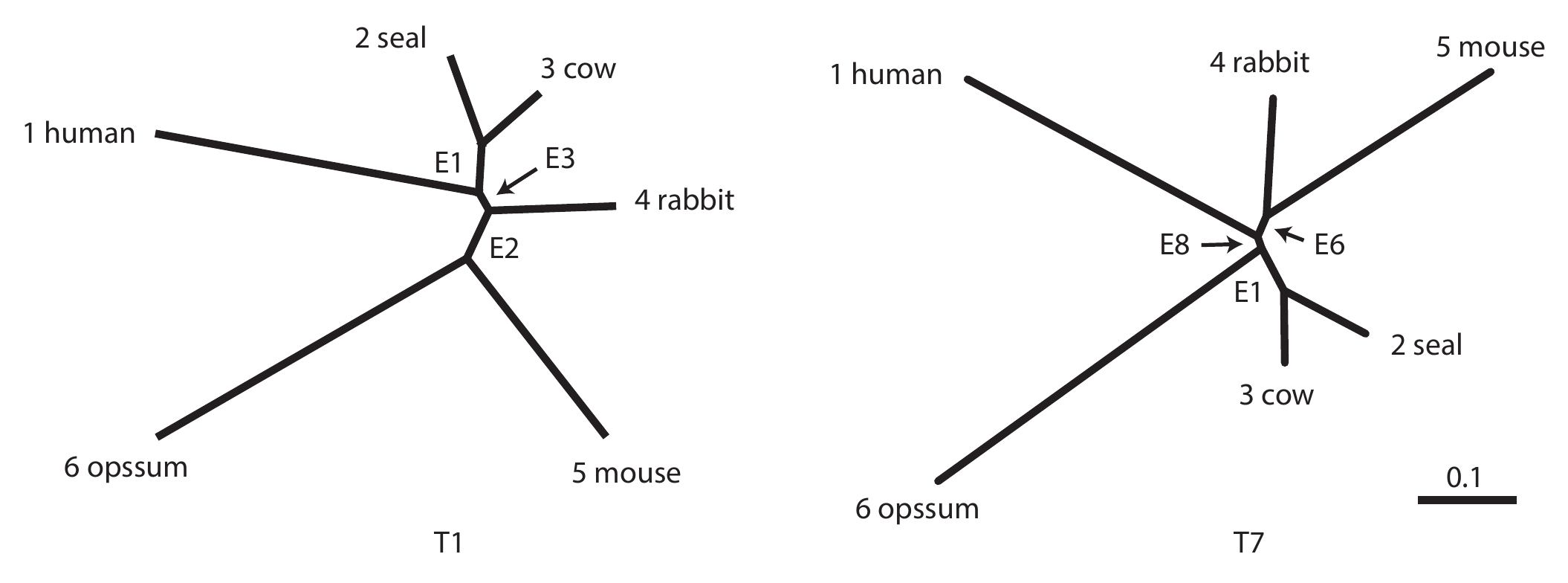}
\end{center}
\caption{Examples of two unrooted trees T1 and T7.
Branch lengths represent ML estimates of parameters (expected number of substitutions per site).
T1 includes edges E1, E2 and E3, and T7 includes E1, E6 and E8.}\label{fig:mam-trees}
\end{figure}

The number of unrooted trees for six taxa is 105. These trees are reordered by their likelihood values and labelled as T1, T2, $\ldots$, T105.  T1 is the ML tree  as shown in figure~\ref{fig:mam-trees} and its tree topology is represented as (((1(23))4)56).
There are three internal branches (we call them as edges) in T1, which are labelled as E1, E2 and E3.
For example, E1 splits the six taxa as $\{23|1456\}$ and the partition of six taxa is represented as \texttt{-++---},
where \texttt{+}/\texttt{-} indicates taxa $1,\ldots,6$ from left to right and \texttt{++} indicates the clade $\{23\}$
  (we set \texttt{-} for taxon 6, since it is treated as the outgroup).
There are 25 edges in total, and each tree is specified by selecting three edges from them, although not all the combinations of three edges are allowed.

The result of phylogenetic analysis is summarized in table~\ref{tab:mam-trees} for trees and table~\ref{tab:mam-edges} for edges.
Three types of $p$-values are computed for each tree as well as for each edge.
BP is the bootstrap probability \citep{Felsenstein:1985:CLP} and AU is the approximately unbiased $p$-value \citep{shimodaira2002approximately}. 
Bootstrap probabilities are computed by the non-parametric bootstrap resampling~\citep{Efron:1979:BMA} described in Section~\ref{sec:rell}.
The theory and the algorithm of BP and AU will be reviewed in Section~\ref{sec:theory}.
Since we are testing many trees and edges at the same time, there is potentially a danger of selection bias.
The issue of selection bias has been discussed in \cite{shimodaira1999multiple} for introducing the method of multiple comparisons of log-likelihoods (also known as \emph{Shimodaira-Hasegawa test}) and in \cite{shimodaira2002approximately} for introducing AU test.
However, these conventional methods are only taking care of the multiplicity of comparing many log-likelihood values for computing just one $p$-value instead of many $p$-values at the same time.
Therefore, we intend to further adjust the AU $p$-value by introducing the selective inference $p$-value, denoted as SI.
The theory and the algorithm of SI will be explained in Section~\ref{sec:selective_inference} based on the geometric theory given in Section~\ref{sec:theory}.
After presenting the methods, we will revisit the phyloegnetic inference in Section~\ref{sec:si_phylo}.

\begin{figure}[h!]
\begin{center}
\includegraphics[width=15cm]{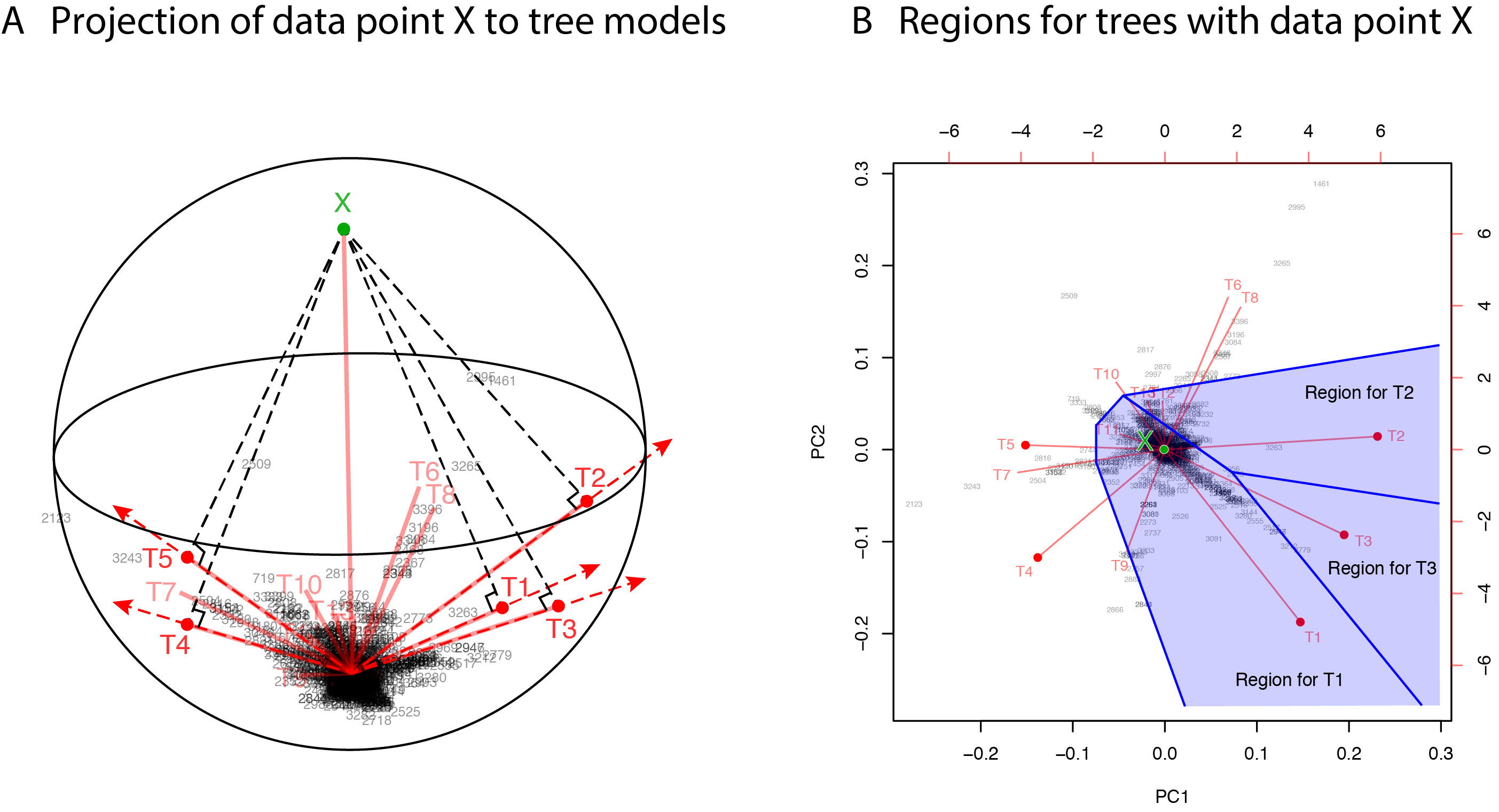}
\end{center}
\caption{Model map: Visualization of ML estimates of probability distributions for the best 15 trees. 
The origin represents the star-shaped tree topology (obtained by reducing the internal branches to zero length).
Sites of amino acid sequences $t=1,\ldots,n$ (black numbers) and probability distributions for trees T1$,\ldots,$T15 (red segments) are drawn by biplot of PCA. Auxiliary lines are drawn by hand.
(A)~3-dimensional visualization using PC1, PC2 and PC3. The reconstructed data point $X$ is also shown (green point).
The ML estimates are represented as the end points of the red segments (shown by red points only for the best five trees),
and they are placed on the sphere with the origin and $X$ being placed at the poles.
(B)~The top-view of model map.
Regions for the best three trees T$i$, $i=1,2,3$ (blue shaded regions) are illustrated; T$i$ will be the ML tree if $X$ is included in the region for T$i$.
}\label{fig:mam-models}
\end{figure}

For developing the geometric theory in Sections~\ref{sec:theory} and \ref{sec:selective_inference},
we formulate tree selection as a mathematical formulation known as \emph{the problem of regions}~\citep{Efron:Halloran:Holmes:1996:BCL,Efron:Tibshirani:1998:PR}.
For better understanding the geometric nature of the theory, the problem of regions is explained below for phylogenetic inference, although the algorithm is simple enough to be implemented without understanding the theory.
Considering the space of probability distributions~\citep{amari2007methods}, the parametric models for trees are represented as manifolds in the space. The dataset (or the empirical distribution) can also be represented as a ``data point'' $X$ in the space, and the ML estimates for trees are represented as projections to the manifolds.
This is illustrated in the visualization of probability distributions of figure~\ref{fig:mam-models}A using log-likelihood vectors of models \citep{shimodaira2001multiple}, where models are simply indicated as red lines from the origin; see Section~\ref{sec:model_map} for details.
This visualization may be called as \emph{model map}.
The point $X$ is actually reconstructed as the minimum full model containing all the trees as submodels,
and the Kullback-Leibler divergence between probability distributions is represented as the squared distance between points; see eq.~(\ref{eq:var-divergence}).
Computation of $X$ is analogous to the Bayesian model averaging, but based on the ML method.
For each tree, we can think of a region in the space so that this tree becomes the ML tree when $X$ is included in the region.
The regions for T1, T2 and T3 are illustrated in figure~\ref{fig:mam-models}B, and the region for E2 is the union of these three regions.

In figure~\ref{fig:mam-models}A, $X$ is very far from any of the tree models, suggesting that all the models are wrong;
the likelihood ratio statistic for testing T1 against the full model is 113.4, which is highly significant as $\chi^2_8$~\citep[Section~5]{shimodaira2001multiple}.
Instead of testing whether tree models are correct or not, we test whether models are significantly better than the others.
As seen in figure~\ref{fig:mam-models}B, $X$ is in the region for T1, meaning that the model for T1 is better than those for the other trees.
For convenience,  observing $X$ in the region for T1, we state that T1 is \emph{supported} by the data.
Similarly, $X$ is in the region for E2 that consists of the three regions for T1, T2, T3, thus indicating that E2 is \emph{supported} by the data.
Although T1 and E2 are supported by the data, there is still uncertainty as to whether the true evolutionary history of lineages is depicted because the location of $X$ fluctuates randomly.
Therefore, statistical confidence of the outcome needs to be assessed.  A mathematical procedure for statistically evaluating the outcome is provided in the following sections.

\section{Non-Selective Inference for the Problem of Regions} \label{sec:theory}

\subsection{The Problem of Regions} \label{sec:geometry}
For developing the theory, we consider $(m+1)$-dimensional multivariate normal random vector $\bm{Y},$ $m
\ge0$,
with unknown mean vector $\bm{\mu}\in\mathbb{R}^{m+1}$ and the identity variance matrix $\bm{I}_{m+1}$:
\begin{equation} \label{eq:y-model}
    \bm{Y} \sim N_{m+1}(\bm{\mu}, \bm{I}_{m+1}).
\end{equation}
A region of interest such as tree and edge is denoted as $\mathcal{R}\subset \mathbb{R}^{m+1}$, and its complement set is denoted as $\mathcal{R}^C =  \mathbb{R}^{m+1} \setminus \mathcal{R}$.
There are $K_\text{all}$ regions $\mathcal{R}_i$, $i=1,\ldots, K_\text{all}$, and we simply write $\mathcal{R}$ for one of them by dropping the index $i$.
Observing $\bm{Y}=\bm{y}$, the null hypothesis $H_0: \bm{\mu}\in\mathcal{R}$ is tested against the alternative hypothesis $H_1: \bm{\mu}\in\mathcal{R}^C$.
This setting is called \emph{problem of regions}, and the geometric theory for non-selective inference for slightly generalized settings (e.g., exponential family of distributions) has been discussed in \cite{Efron:Tibshirani:1998:PR,shimodaira2004approximately}.
This theory allows arbitrary shape of $\mathcal{R}$ without assuming a particular shape such as half-space or sphere, and only requires the expression (\ref{eq:region-uv}) of Section~\ref{sec:smooth_surface}.

\begin{figure}[h!]
\begin{center}
\includegraphics[width=10cm]{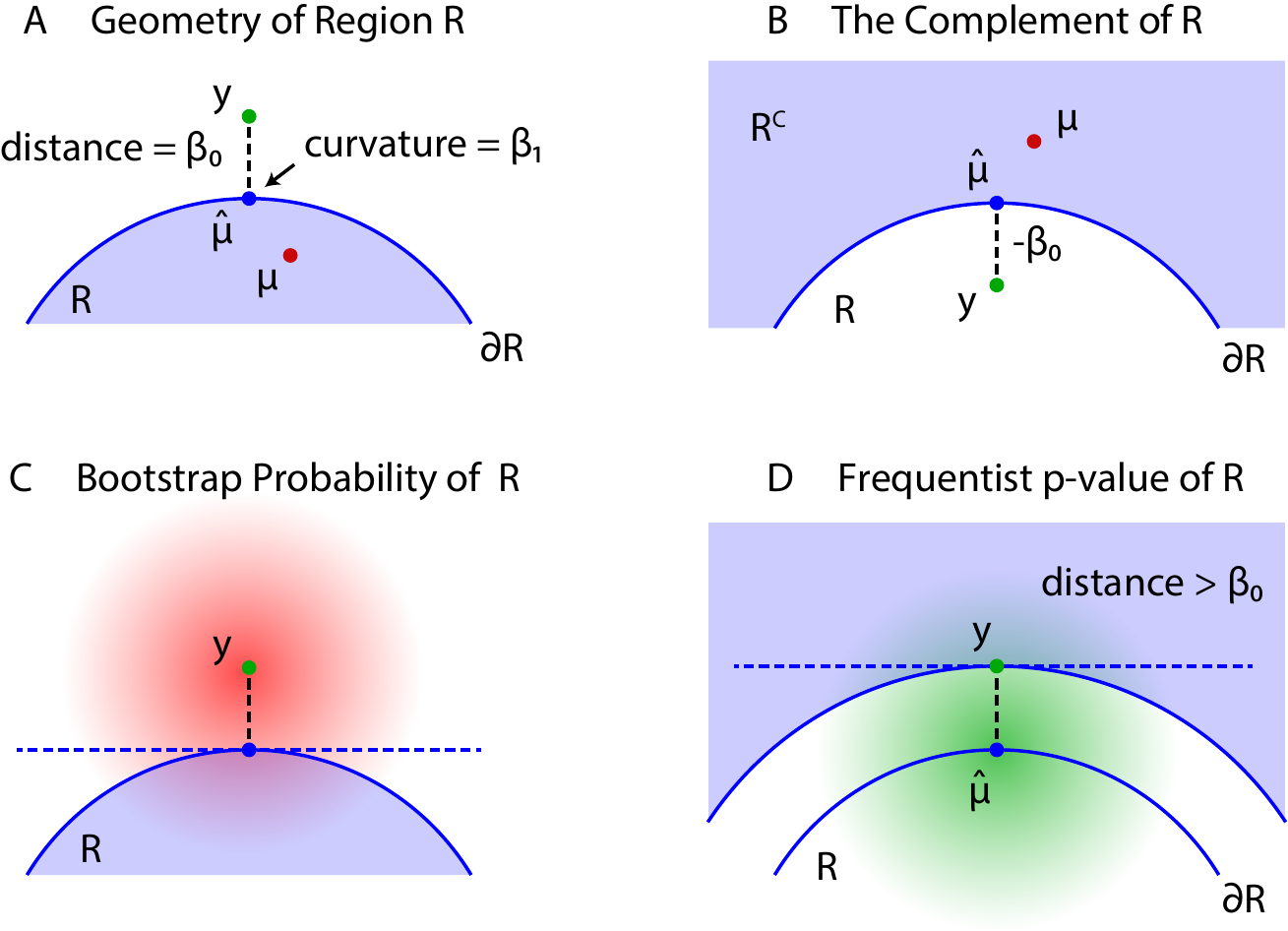}
\end{center}
\caption{Problem of regions.
(A)~$\beta_0>0$ when $\bm{y}\in \mathcal{R}^C$, then select the null hypothesis $\bm{\mu}\in\mathcal{R}$.
(B)~$\beta_0\le 0$ when $\bm{y}\in\mathcal{R}$, then select the null hypothesis $\bm{\mu}\in\mathcal{R}^C$.
(C)~The bootstrap distribution of $\bm{Y}^* \sim N_{m+1}(\bm{y},\bm{I}_{m+1})$ (red shaded distribution).
(D)~The null distribution of $\bm{Y} \sim N_{m+1}(\bm{\hat\mu},\bm{I}_{m+1}) $ (green shaded distribution).
}
\label{fig:geometry}
\end{figure}

The problem of regions is well described by geometric quantities (figure~\ref{fig:geometry}). Let $\bm{\hat\mu}$ be the projection of $\bm{y}$ to the boundary surface $\partial \mathcal{R}$ defined as
\begin{equation*}
    \bm{\hat\mu} = \argmin_{\bm{\mu} \in \partial \mathcal{R}} \| \bm{y} - \bm{\mu} \|,
\end{equation*}
and $\beta_0$ be the \emph{signed distance} defined as $\beta_0 = \| \bm{y} - \bm{\hat\mu} \| > 0$ for $\bm{y}\in\mathcal{R}^C$ and $\beta_0 = -\| \bm{y} - \bm{\hat\mu} \| \le 0$ for $\bm{y}\in\mathcal{R}$; see figures~\ref{fig:geometry}A and \ref{fig:geometry}B, respectively.
A large $\beta_0$ indicates the evidence for rejecting $H_0:\bm{\mu}\in\mathcal{R}$, but computation of $p$-value will also depend on the shape of $\mathcal{R}$.  
There should be many parameters for defining the shape, but we only need the \emph{mean curvature} of $\partial \mathcal{R}$ at $\bm{\hat\mu}$, which represents the amount of surface bending. It is denoted as $\beta_1\in\mathbb{R}$, and defined in (\ref{eq:beta1}).

Geometric quantities $\beta_0$ and $\beta_1$ of regions for trees (T1$,\ldots,$ T105) and edges (E1$,\ldots,$ E25) are plotted in figure~\ref{fig:mam-beta}, and these values are also found in tables~\ref{tab:mam-trees} and \ref{tab:mam-edges}.
Although the phylogenetic model of evolution for the molecular dataset $\mathcal{X}_n = (\bm{x}_1,\ldots,\bm{x}_n)$ is different from the multivariate normal model  (\ref{eq:y-model}) for $\bm{y}$, the multiscale bootstrap method of Section~\ref{sec:multiscale_bootstrap} estimates $\beta_0$ and $\beta_1$ using the non-parametric bootstrap probabilities (Section~\ref{sec:rell}) with bootstrap replicates $\mathcal{X}^*_{n'}$ for several values of sample size $n'$.

\subsection{Bootstrap Probability} \label{sec:bootstrap_probability}

For simulating (\ref{eq:y-model}) from $\bm{y}$, we may generate replicates $\bm{Y}^*$ from the bootstrap distribution
(figure~\ref{fig:geometry}C)
\begin{equation} \label{eq:y-boot}
    \bm{Y}^* \sim N_{m+1}(\bm{y}, \bm{I}_{m+1}),
\end{equation}
and define bootstrap probability (BP) of $\mathcal{R}$ as the probability of $\bm{Y}^*$ being included in the region $\mathcal{R}$:
\begin{equation} \label{eq:bp-def}
    \text{BP}(\mathcal{R} | \bm{y}) := P(\bm{Y}^* \in \mathcal{R} | \bm{y}).
\end{equation}
$\text{BP}(\mathcal{R} | \bm{y})$ can be interpreted as the Bayesian posterior probability $P(\bm{\mu}\in\mathcal{R}| \bm{y})$,
because, by assuming the flat prior distribution $\pi(\bm{\mu}) = $ constant, the posterior distribution $\bm{\mu} | \bm{y} \sim N_{m+1}(\bm{y}, \bm{I}_{m+1})$ is identical to the distribution of $\bm{Y}^*$ in (\ref{eq:y-boot}).
An interesting consequence of the geometric theory of \cite{Efron:Tibshirani:1998:PR} is that
BP can be expressed as
\begin{equation} \label{eq:bp-beta}
    \text{BP}(\mathcal{R} | \bm{y}) \simeq \bar\Phi(\beta_0 + \beta_1),
\end{equation}
where $\simeq$ indicates the \emph{second order asymptotic accuracy}, meaning that the equality is correct up to $O_p(n^{-1/2})$ with error of order $O_p(n^{-1})$; see Section~\ref{sec:smooth_surface}.

For understanding the formula (\ref{eq:bp-beta}), assume that $\mathcal{R}$ is a half space so that $\partial \mathcal{R}$ is flat and $\beta_1 = 0$. Since we only have to look at the axis orthogonal to $\partial \mathcal{R}$, the distribution of signed distance is identified as (\ref{eq:z-normal}) with $\beta_0=z$.
The bootstrap distribution for (\ref{eq:z-normal}) is $Z^* \sim N(z,1)$, and bootstrap probability is expressed as $P(Z^* \le 0 | z)=\bar\Phi(z)$.
Therefore, we have $\text{BP}(\mathcal{R} | \bm{y})  =  \bar\Phi(\beta_0)$.
For general $\mathcal{R}$ with curved $\partial \mathcal{R}$, the formula (\ref{eq:bp-beta}) adjusts the bias caused by $\beta_1$. As seen in figure~\ref{fig:geometry}C, $\mathcal{R}$ becomes smaller for $\beta_1>0$ than $\beta_1=0$, and BP becomes smaller. 

BP of $\mathcal{R}^C$ is closely related to BP of  $\mathcal{R}$. From the definition,
\begin{equation} \label{eq:bp-beta-c}
   \text{BP}(\mathcal{R}^C | \bm{y}) =  1 - \text{BP}(\mathcal{R} | \bm{y}) \simeq 
   1-\bar\Phi(\beta_0 + \beta_1) =  \bar\Phi(-\beta_0 - \beta_1).
\end{equation}
The last expression also implies that the signed distance and the mean curvature of $\mathcal{R}^C$ is $-\beta_0$ and $-\beta_1$, respectively; this relation is also obtained by reversing the sign of $v$ in (\ref{eq:region-uv}).

\subsection{Approximately Unbiased Test} \label{sec:autest}

Although $\text{BP}(\mathcal{R}|\bm{y})$ may work as a Bayesian confidence measure, we would like to have a frequentist confidence measure for testing $H_0: \bm{\mu}\in\mathcal{R}$ against  $H_1: \bm{\mu}\in\mathcal{R}^C$.
The signed distance of $\bm{Y}$ is denoted as $\beta_0(\bm{Y})$,
and consider the region $\{\bm{Y} \mid \beta_0(\bm{Y}) > \beta_0 \}$ in which
the signed distance is larger than the observed value $\beta_0 = \beta_0(\bm{y})$.
Similar to (\ref{eq:z-au}), we then define an approximately unbiased (AU) $p$-value as
\begin{equation} \label{eq:au-def}
    \text{AU}(\mathcal{R}|\bm{y}) := P( \beta_0(\bm{Y}) > \beta_0  \mid \bm{\mu} = \bm{\hat\mu} )
     = \text{BP}( \{\bm{Y} \mid \beta_0(\bm{Y}) > \beta_0 \} | \bm{\hat\mu}),
\end{equation}
where the probability is calculated for $\bm{Y} \sim N_{m+1}(\bm{\hat\mu} , \bm{I}_{m+1} )$ as illustrated in figure~\ref{fig:geometry}D.
The shape of the region $ \{ \bm{Y} \mid  \beta_0(\bm{Y}) > \beta_0 \} $ is very similar to the shape of $\mathcal{R}^C$; the difference is in fact only $O_p(n^{-1})$. 
Let us think of a point $\bm{y}'$ with signed distance $-\beta_0$ (shown as $\bm{y}$ in figure~\ref{fig:geometry}B).
Then we have
\begin{equation} \label{eq:au-beta}
    \text{AU}(\mathcal{R}|\bm{y}) \simeq \text{BP}( \mathcal{R}^C | \bm{y}') \simeq \bar\Phi(\beta_0 - \beta_1),
\end{equation}
where the last expression is obtained by substituting $(-\beta_0,\beta_1)$ for $(\beta_0, \beta_1)$ in (\ref{eq:bp-beta-c}).
This formula computes AU from $(\beta_0,\beta_1)$.
An intuitive interpretation of (\ref{eq:au-beta}) is explained in Section~\ref{sec:si_simple_case}.

In non-selective inference, $p$-values are computed using formula (\ref{eq:au-beta}).  If $\text{AU}(\mathcal{R}|\bm{y}) < \alpha$, the null hypothesis $H_0: \bm{\mu}\in\mathcal{R}$ is rejected and the alternative hypothesis $H_1: \bm{\mu}\in\mathcal{R}^C$ is accepted.
This test procedure is approximately unbiased, because it controls the non-selective type-I error as
\begin{equation} \label{eq:au-typeierror}
    P\bigl(     \text{AU}(\mathcal{R}|\bm{Y})  < \alpha \mid  \bm{\mu} \in \partial \mathcal{R}  \bigr) \simeq \alpha,
\end{equation}
and the rejection probability increases as $\bm{\mu}$ moves away from $\mathcal{R}$, while
it decreases as $\bm{\mu}$ moves into $\mathcal{R}$.

Exchanging the roles of $\mathcal{R}$ and $\mathcal{R}^C$ also allows for another hypothesis testing.
AU of $\mathcal{R}^C$ is obtained from (\ref{eq:au-def}) by reversing the inequality
as $\text{AU}(\mathcal{R}^C | \bm{y}) =  \text{BP}( \{\bm{Y} \mid \beta_0(\bm{Y}) < \beta_0 \} | \bm{\hat\mu}) = 1 - \text{AU}(\mathcal{R} | \bm{y})$.
This is also confirmed by substituting $(-\beta_0, -\beta_1)$, i.e., the geometric quantities of $\mathcal{R}^C$,  for $(\beta_0, \beta_1)$ in (\ref{eq:au-beta}) as
\begin{equation} \label{eq:au-beta-c}
\text{AU}(\mathcal{R}^C | \bm{y}) \simeq \bar\Phi(-\beta_0 + \beta_1) \simeq 1 - \text{AU}(\mathcal{R} | \bm{y}).
\end{equation}
If $\text{AU}(\mathcal{R}^C|\bm{y}) < \alpha$ or equivalently $\text{AU}(\mathcal{R}|\bm{y}) > 1- \alpha$, then we reject $H_0: \bm{\mu}\in\mathcal{R}^C$ and accept $H_1: \bm{\mu}\in\mathcal{R}$.

\subsection{Multiscale Bootstrap} \label{sec:multiscale_bootstrap}

In order to estimate $\beta_0$ and $\beta_1$ from bootstrap probabilities, we consider a generalization of (\ref{eq:y-boot}) as
\begin{equation} \label{eq:y-msboot}
    \bm{Y}^* \sim N_{m+1}(\bm{y}, \sigma^2\bm{I}_{m+1}),
\end{equation}
for a variance $\sigma^2>0$, and define multiscale bootstrap probability of $\mathcal{R}$ as
\begin{equation} \label{eq:bps-def}
    \text{BP}_{\sigma^2}(\mathcal{R} | \bm{y}) := P_{\sigma^2}(\bm{Y}^* \in \mathcal{R} | \bm{y}),
\end{equation}
where $P_{\sigma^2}$ indicates the probability with respect to (\ref{eq:y-msboot}).

Although our theory is based on the multivariate normal model, the actual implementation of the algorithm uses the non-parametric bootstrap probabilities in Section~\ref{sec:rell}.
To fill the gap between the two models, we consider a non-linear transformation $\bm{f}_n$ so that the multivariate normal model holds at least approximately for $\bm{y} = \bm{f}_n(\mathcal{X}_n)$ and  $\bm{Y}^* = \bm{f}_n(\mathcal{X}^*_{n'})$.
An example of $\bm{f}_n$ is given in (\ref{eq:fn-tree}) for phylogenetic inference.
Surprisingly, a specification of $\bm{f}_n$ is \emph{not required} for computing $p$-values,  but we simply assume the existence of such a transformation; this property may be called as ``bootstrap trick''.
For phylogenetic inference, we compute the non-parametric bootstrap probabilities by (\ref{eq:raw-bp-tree}) and substitute these values for (\ref{eq:bps-def}) with $\sigma^2=n/n'$.

For estimating $\beta_0$ and $\beta_1$, we need to have a scaling law which explains how $\text{BP}_{\sigma^2}$ depends on the scale $\sigma$.
We rescale (\ref{eq:y-msboot}) by multiplying $\sigma^{-1}$ so that $ \sigma^{-1} \bm{Y}^* \sim N_{m+1}(\sigma^{-1} \bm{y}, \bm{I}_{m+1})$ has the variance $\sigma^2=1$.
$\bm{y}$ and $\mathcal{R}$ are now resaled by the factor $\sigma^{-1}$, which amounts to signed distance $\beta_0 \sigma^{-1}$ and mean curvature $\beta_1 \sigma$~\citep{shimodaira2004approximately}.
Therefore, by substituting $(\beta_0 \sigma^{-1}, \beta_1 \sigma)$ for $(\beta_0, \beta_1)$ in (\ref{eq:bp-beta}), we obtain
\begin{equation} \label{eq:bps-beta}
    \text{BP}_{\sigma^2}(\mathcal{R} | \bm{y}) \simeq \bar\Phi(\beta_0\sigma^{-1} + \beta_1\sigma).
\end{equation}
For better illustrating how $ \text{BP}_{\sigma^2}$ depends on $\sigma^2$, we define
\begin{equation} \label{eq:psi-beta}
    \psi_{\sigma^2}(\mathcal{R} | \bm{y}) := \sigma \bar\Phi^{-1} (  \text{BP}_{\sigma^2}(\mathcal{R} | \bm{y}) )
    \simeq \beta_0 + \beta_1 \sigma^2.
\end{equation}
We can estimate $\beta_0$ and $\beta_1$ as regression coefficients by fitting the linear model (\ref{eq:psi-beta}) in terms of $\sigma^2$ to the observed values of non-parametric bootstrap probabilities (figure~\ref{fig:mam-fitting}).
Interestingly, (\ref{eq:au-beta}) is rewritten as $\text{AU}(\mathcal{R} | \bm{y}) \simeq \bar\Phi(\psi_{-1}(\mathcal{R} | \bm{y}) ) $ by formally letting $\sigma^2= -1$ in the last expression of (\ref{eq:psi-beta}), meaning that AU corresponds to $n'=-n$.
Although $\sigma^2$ should be positive in (\ref{eq:bps-beta}), we can think of negative $\sigma^2$ in $\beta_0 + \beta_1 \sigma^2$.
See Section~\ref{sec:msboot-detail} for details of model fitting and extrapolation to negative $\sigma^2$.

\section{Selective Inference for the Problem of Regions} \label{sec:selective_inference}

\subsection{Approximately Unbiased Test for Selective Inference} \label{sec:si_geometry}

In order to argue selective inference for the problem of regions, we have to specify the selection event.
Let us consider a selective region $\mathcal{S}\subset \mathcal{R}^{m+1}$ so that we perform the hypothesis testing only when $\bm{y}\in\mathcal{S}$. \cite{terada2017selective} considered a general shape of $\mathcal{S}$, but here we treat only two special cases of $\mathcal{S} = \mathcal{R}^C$ and $\mathcal{S} = \mathcal{R}$; see Section~\ref{sec:si_general}.
Our problem is formulated as follows.
Observing $\bm{Y}=\bm{y}$ from the multivariate normal model (\ref{eq:y-model}), we first check whether $\bm{y}\in \mathcal{R}^C$ or  $\bm{y}\in \mathcal{R}$.
If $\bm{y}\in \mathcal{R}^C$ and we are interested in the null hypothesis $H_0: \bm{\mu}\in\mathcal{R}$, then we may test it against the alternative hypothesis $H_1: \bm{\mu}\in\mathcal{R}^C$.
If $\bm{y}\in \mathcal{R}$ and we are interested in the null hypothesis $H_0: \bm{\mu}\in\mathcal{R}^C$, then we may test it against the alternative hypothesis $H_1: \bm{\mu}\in\mathcal{R}$.
In this paper,
the former case ($\bm{y}\in\mathcal{R}^C$, and so $\beta_0>0$) is called as \emph{outside mode}, and the latter case ($\bm{y}\in\mathcal{R}$, and so $\beta_0\le0$) is called as \emph{inside mode}.
We do not know which of the two modes of testing is performed until we observe $\bm{y}$. 

Let us consider the outside mode by assuming that $\bm{y}\in \mathcal{R}^C$, where $\beta_0 > 0$.
Recalling that $p(z,c)=p(z)/\bar\Phi(c)$ in Section~\ref{sec:intro}, we divide
$\text{AU}(\mathcal{R}|\bm{y})$ by the selection probability to define a selective inference $p$-value as
\begin{equation} \label{eq:si-def}
    \text{SI} (\mathcal{R} | \bm{y}) :=
    \frac{P( \beta_0(\bm{Y}) > \beta_0  \mid \bm{\mu} = \bm{\hat\mu} )}{P(\bm{Y}\in \mathcal{R}^C \mid \bm{\mu} = \bm{\hat\mu} )}
    = \frac{ \text{AU}(\mathcal{R}|\bm{y})}{\text{BP}( \mathcal{R}^C | \bm{\hat\mu})}.
\end{equation}
From the definition, $\text{SI} (\mathcal{R} | \bm{y}) \in (0,1)$, because $ \{\bm{Y} \mid  \beta_0(\bm{Y}) > \beta_0   \} \subset\mathcal{R}^C $ for $\beta_0>0$.
This $p$-value is computed from $(\beta_0,\beta_1)$ by
\begin{equation} \label{eq:si-beta}
    \text{SI} (\mathcal{R} | \bm{y}) \simeq \frac{\bar\Phi(\beta_0 - \beta_1)}{\bar\Phi(-\beta_1)},
\end{equation}
where $\text{BP}( \mathcal{R}^C | \bm{\hat\mu}) = \bar\Phi(-\beta_1)$ is obtained by substituting $(0,\beta_1)$ for $(\beta_0, \beta_1)$ in (\ref{eq:bp-beta-c}).
An intuitive justification of (\ref{eq:si-beta}) is explained in Section~\ref{sec:si_simple_case}.

For the outside mode of selective inference, $p$-values are computed using formula (\ref{eq:si-beta}). If $\text{SI}(\mathcal{R}|\bm{y}) < \alpha$, then 
reject $H_0: \bm{\mu}\in\mathcal{R}$ and accept $H_1: \bm{\mu}\in\mathcal{R}^C$.
This test procedure is approximately unbiased, because it controls the selective type-I error as
\begin{equation} \label{eq:si-typeierror}
    P\bigl(     \text{SI}(\mathcal{R}|\bm{Y})  < \alpha \mid  \bm{Y}\in \mathcal{R}^C, \bm{\mu} \in \partial \mathcal{R}  \bigr) \simeq \alpha,
\end{equation}
and the rejection probability increases as $\bm{\mu}$ moves away from $\mathcal{R}$, while
it decreases as $\bm{\mu}$ moves into $\mathcal{R}$.

Now we consider the inside mode by assuming that $\bm{y}\in \mathcal{R}$, where $\beta_0 \le 0$.
SI of $\mathcal{R}^C$ is obtained from (\ref{eq:si-def}) by exchanging the roles of $\mathcal{R}$ and $\mathcal{R}^C$.
\begin{equation} \label{eq:si-def-beta-c}
    \text{SI} (\mathcal{R}^C | \bm{y})
    = \frac{ \text{AU}(\mathcal{R^C}|\bm{y})}{\text{BP}( \mathcal{R} | \bm{\hat\mu})}
    \simeq \frac{\bar\Phi(-\beta_0 + \beta_1)}{\bar\Phi(\beta_1)}.
\end{equation}
For the inside mode of selective inference, $p$-values are computed using formula (\ref{eq:si-def-beta-c}).
If $\text{SI}(\mathcal{R}^C|\bm{y}) < \alpha$, then reject $H_0: \bm{\mu}\in\mathcal{R}^C$ and accept $H_1: \bm{\mu}\in\mathcal{R}$.
Unlike the non-selective $p$-value $\text{AU}(\mathcal{R}^C|\bm{y})$,  
$\text{SI}(\mathcal{R}^C|\bm{y}) < \alpha$ is \emph{not} equivalent to  $\text{SI}(\mathcal{R}|\bm{y}) > 1- \alpha$, because $\text{SI}(\mathcal{R}|\bm{y})  + \text{SI}(\mathcal{R}^C|\bm{y}) \neq 1$.
For convenience, we define
\begin{equation} \label{eq:si3}
    \text{SI}'(\mathcal{R}|\bm{y}) := 
    \begin{cases}
    \text{SI}(\mathcal{R}|\bm{y})  & \bm{y}\in\mathcal{R}^C\\
    1-\text{SI}(\mathcal{R}^C|\bm{y})  & \bm{y}\in\mathcal{R}
    \end{cases}
\end{equation}
so that $\text{SI}' > 1-\alpha$ implies $\text{SI}(\mathcal{R}^C|\bm{y}) <\alpha$.
In our numerical examples of figure~\ref{fig:mam-beta}, tables~\ref{tab:mam-trees} and \ref{tab:mam-edges},
$\text{SI}'$ is simply denoted as SI.
We do not need to consider (\ref{eq:si3}) for BP and AU, because
 $\text{BP}'(\mathcal{R}|\bm{y}) = \text{BP}(\mathcal{R}|\bm{y})$ and $\text{AU}'(\mathcal{R}|\bm{y}) = \text{AU}(\mathcal{R}|\bm{y})$ from (\ref{eq:bp-beta-c}) and (\ref{eq:au-beta-c}).


\subsection{Shortcut Computation of SI}

We can compute SI from BP and AU.
This will be useful for reanalyzing the results of previously published researches.
Let us write $\text{BP} = \text{BP}(\mathcal{R}|\bm{y})$ and $\text{AU} = \text{AU}(\mathcal{R}|\bm{y})$.
From (\ref{eq:bp-beta}) and (\ref{eq:au-beta}), we have
\begin{align*}
    \beta_0 &= \tfrac{1}{2} \Bigl(  \bar\Phi^{-1}(\text{BP}) + \bar\Phi^{-1}(\text{AU})  \Bigr)\\
    \beta_1 &= \tfrac{1}{2} \Bigl(  \bar\Phi^{-1}(\text{BP}) - \bar\Phi^{-1}(\text{AU})  \Bigr).
\end{align*}
We can compute SI from $\beta_0$ and $\beta_1$ by (\ref{eq:si-beta}) or (\ref{eq:si-def-beta-c}).
More directly, we may compute
\begin{align*} 
    \text{SI}(\mathcal{R}|\bm{y}) &= \frac{\text{AU}}{\bar\Phi\Bigl\{ \tfrac{1}{2} \Bigl(  \bar\Phi^{-1}(\text{AU}) - \bar\Phi^{-1}(\text{BP})  \Bigr)  \Bigr\}}  \\
    \text{SI}(\mathcal{R}^C|\bm{y}) &= \frac{1-\text{AU}}{\bar\Phi\Bigl\{ \tfrac{1}{2} \Bigl(  \bar\Phi^{-1}(\text{BP}) - \bar\Phi^{-1}(\text{AU})  \Bigr)  \Bigr\}}.
\end{align*}


\subsection{Revisiting the Phylogenetic Inference} \label{sec:si_phylo}

In this section, the analytical procedure outlined in Section~\ref{sec:phylo} is used to determine relationships among human, mouse, and rabbit.
The question is: Which of mouse or human is closer to rabbit?
The traditional view \citep{novacek1992mammalian} is actually supporting E6, the clade of rabbit and mouse, which is consistent with T4, T5 and T7.
Based on molecular analysis, \cite{graur1996phylogenetic} strongly suggested that rabbit is closer to human than mouse, thus supporting E2, which is consistent with T1, T2 and T3.
However, \cite{halanych1998lagomorphs} criticized it by pointing out that E2 is an artifact caused by the \emph{long branch attraction} (LBA) between mouse and opossum.
In addition, \cite{shimodaira1999multiple,shimodaira2002approximately} suggested that T7 is not rejected by multiplicity adjusted tests.
\cite{shimodaira2005assessing} showed that T7 becomes the ML tree by resolving the LBA using a larger dataset with more taxa.
Although T1 is the ML tree based on the dataset with fewer taxa, T7 is presumably the true tree as indicated by later researches.
With these observations in mind, we retrospectively interpret $p$-values in tables~\ref{tab:mam-trees} and \ref{tab:mam-edges}.

The results are shown below for the two test modes (inside and outside) as defined in Section~\ref{sec:si_geometry}.
The extent of multiplicity and selection bias depends on the number of regions under consideration, thus these numbers are considered for interpreting the results.
The numbers of regions related to trees and edges are summarized in table~\ref{tab:Knumbers}; see Section~\ref{sec:number_regions} for details.

In inside mode, the null hypothesis $H_0: \bm{\mu}\in \mathcal{R}_i^C$ is tested against the alternative hypothesis $H_1: \bm{\mu}\in \mathcal{R}_i$ for $\bm{y}\in\mathcal{R}_i$ (i.e., $\beta_0\le0$). This applies to the regions for T1, E1, E2 and E3, and they are \emph{supported} by the data in the sense mentioned in the last paragraph of Section~\ref{sec:phylo}.
When $H_0$ is rejected by a test procedure, it is claimed that $\mathcal{R}_i$ is  \emph{significantly supported} by the data, indicating $H_1$ holds true.
For convenience, the null hypothesis $H_0$ is said like E1 is not true, and the alternative hypothesis $H_1$ is said like E1 is true; then rejection of $H_0$ implies that E1 is true.
This procedure looks unusual, but makes sense when both $\mathcal{R}_i$ and $\mathcal{R}_i^C$ are regions with nonzero volume.
Note that selection bias can be very large in the sense that $K_\text{select}/K_\text{all}\approx 0$ for many taxa,
and non-selective tests may lead to many false positives because $K_\text{true}/K_\text{all}\approx 1$.
Therefore selective inference should be used in inside mode.

In outside mode, the null hypothesis $H_0: \bm{\mu}\in \mathcal{R}_i$ is tested against the alternative hypothesis $H_1: \bm{\mu}\in \mathcal{R}_i^C$ for $\bm{y}\in\mathcal{R}_i^C$ (i.e., $\beta_0>0$). This applies to the regions for T2, ..., T105, and E4, ..., E25, and they are \emph{not supported} by the data.
When $H_0$ is rejected by a test procedure, it is claimed that $\mathcal{R}_i$ is rejected.
For convenience, the null hypothesis is said like T9 is true, and the alternative hypothesis is said like T9 is not true;  rejection of $H_0$ implies that T9 is not true.
This is more or less a typical test procedure.
Note that selection bias is minor in the sense that $K_\text{select}/K_\text{all}\approx 1$ for many taxa,
and non-selective tests may result in few false positives because $K_\text{true}/K_\text{all}\approx 0$.
Therefore selective inference is not much beneficial in outside mode.

In addition to $p$-values for some trees and edges, estimated geometric quantities are also shown in the tables.
We confirm that the sign of $\beta_0$ is estimated correctly for all the trees and edges.
The estimated $\beta_1$ values are all positive, indicating the regions are convex.
This is not surprising, because the regions are expressed as intersections of half spaces at least locally (figure~\ref{fig:mam-models}B).

\begin{figure}[h!]
\begin{center}
\includegraphics[width=10cm]{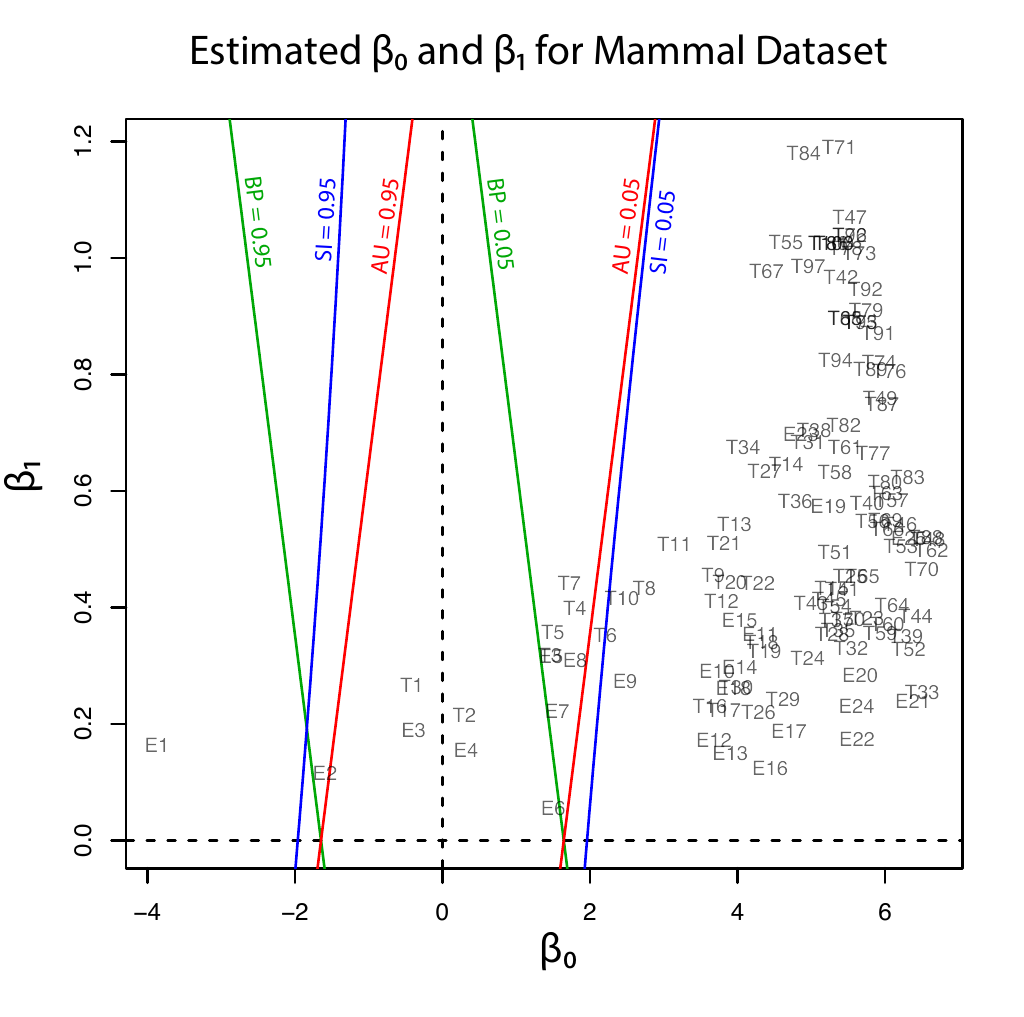}
\end{center}
\caption{Geometric quantities of regions ($\beta_0$ and $\beta_1$) for trees and edges
are estimated by the multiscale bootstrap method (Section~\ref{sec:multiscale_bootstrap}).
The three types of $p$-value (BP, AU, SI) are computed from $\beta_0$ and $\beta_1$,
and their contour lines are drawn at $p=$ 0.05 and 0.95.
}\label{fig:mam-beta}
\end{figure}

Now $p$-values are examined in inside mode.
(T1, E3)~BP, AU, SI are all $p\le0.95$. This indicates that T1 and E3 are \emph{not} significantly supported.
There are nothing claimed to be definite.
(E1)~BP, AU, SI are all $p> 0.95$, indicating E1 is significantly supported.
Since E1 is associated with the best 15 trees T1, ..., T15, some of them are significantly better than the rest of trees T16, ..., T105. 
Significance for edges is common in phylogenetics as well as in hierarchical clustering~\citep{suzuki2006pvclust}.
(E2)~The results split for this presumably wrong edge.
$\text{AU}>0.95$ suggests E2 is significantly supported, whereas $\text{BP},\text{SI} \le 0.95$ are not significant.
AU tends to violate the selective type-I error, leading to false positives or overconfidence in wrong trees/edges, whereas SI is approximately unbiased for the selected hypothesis.
This overconfidence is explained by the inequality $\text{AU}>\text{SI}$ (meant $\text{SI}'$ here) for $\bm{y}\in\mathcal{R}$, which is obtained by comparing (\ref{eq:au-beta-c}) and (\ref{eq:si-def-beta-c}).
Therefore SI is preferable to AU in inside mode.
BP is safer than AU in the sense that $\text{BP}<\text{AU}$ for $\beta_1>0$, but BP is not guaranteed for controlling type-I error in a frequentist sense.
The two inequalities ($\text{SI}, \text{BP}<\text{AU}$) are verified as relative positions of the contour lines at $p=0.95$ in figure~\ref{fig:mam-beta}. The three $p$-values can be very different from each other for large $\beta_1$.

Next $p$-values are examined in outside mode. 
(T2, E4, E6)~BP, AU, SI are all $p\ge0.05$.
They are \emph{not} rejected, and there are nothing claimed to be definite.
(T8, T9, ..., T105, E9,..., E25)~BP, AU, SI are all $p<0.05$.
These trees and edges are rejected.
(T7, E8)~The results split for these presumably true tree and edge.
$\text{BP} < 0.05$ suggests T7 and E8 are rejected, whereas $\text{AU}, \text{SI} \ge 0.05$ are not significant. 
AU is approximately unbiased for controlling the type-I error when $H_0$ is specified in advance~\citep{shimodaira2002approximately}.
Since $\text{BP}<\text{AU}$ for $\beta_1>0$, 
BP violates the type-I error, which results in overconfidence in non-rejected wrong trees.
Therefore BP should be avoided in outside mode.
Inequality $\text{AU}<\text{SI}$ can be shown for $\bm{y}\in\mathcal{R}^C$ by comparing (\ref{eq:au-beta}) and (\ref{eq:si-beta}).
Since the null hypothesis $H_0: \bm{\mu}\in \mathcal{R}$ is chosen after looking at $\bm{y}\in\mathcal{R}^C$,
AU is not approximately unbiased for controlling the selective type-I error, whereas SI adjusts this selection bias. 
The two inequalities ($\text{BP}<\text{AU}<\text{SI}$) are verified as relative positions of the contour lines at $p=0.05$ in figure~\ref{fig:mam-beta}. AU and SI behave similarly (Note: $K_\text{select}/K_\text{all} \approx 1$), while BP is very different from AU and SI for large $\beta_1$.
It is arguable which of AU and SI is appropriate: AU is preferable to SI in tree selection ($K_\text{true} = 1$), because the multiplicity of testing is controlled as $\text{FWER} = P(\text{reject any true null}) = P(\text{AU}(\mathcal{R}_\text{true tree} |\bm{Y}) < \alpha \mid \bm{\mu}\in\mathcal{R}_\text{true tree}) \le \alpha$.
The FWER is multiplied by $K_\text{true}\ge1$ for edge selection, and SI does not fix it either.
For testing edges in outside mode, AU may be used for screening purpose with a small $\alpha$ value such as $\alpha/K_\text{true}$.

\section{Conclusion} \label{sec:conclusion}

We have developed a new method for computing selective inference $p$-values from multiscale bootstrap probabilities, and applied this new method to phylogenetics.
It is demonstrated through theory and a real-data analysis that selective inference  $p$-values are in particular useful for testing selected edges (i.e., clades or clusters of species) to claim that they are supported significantly if $p>1-\alpha$.
On the other hand, the previously proposed non-selective version of approximately unbiased $p$-values are still useful for testing candidate trees to claim that they are rejected if $p<\alpha$.
Although we focused on phylogenetics, our general theory of selective inference may be applied to other model selection problems, or more general selection problems.

\section{Remarks} \label{sec:remarks}

\subsection{Bootstrap resampling of log-likelihoods} \label{sec:rell}

Non-parametric bootstrap is often time consuming for recomputing the maximum likelihood (ML) estimates for bootstrap replicates.
\cite{bib:Kish:Miya:Hase:90:MLI} considered the resampling of estimated log-likelihoods (RELL) method for reducing the computation.
Let $\mathcal{X}_n=(\bm{x}_1,\ldots,\bm{x}_n)$ be the dataset of sample size $n$, where $\bm{x}_t$ is the site-pattern of amino acids at site $t$ for $t=1,\ldots, n$. By resampling $\bm{x}_t$ from $\mathcal{X}_n$ with replacement, we obtain a bootstrap replicate
$\mathcal{X}^*_{n'}=(\bm{x}^*_1,\ldots,\bm{x}^*_{n'})$ of sample size $n'$.
Although $n'=n$ for the ordinary bootstrap, we will use several $n'>0$ values for the multiscale bootstrap.
The parametric model of probability distribution for tree T$i$ is $p_i(\bm{x} ; \bm{\theta}_i)$ for $i=1,\ldots,105$,
and the log-likelihood function is $\ell_i(\bm{\theta}_i ; \mathcal{X}_n) = \sum_{t=1}^n \log p_i(\bm{x}_t ; \bm{\theta}_i)$.
Computation of the ML estimate $\bm{\hat\theta}_i = \argmax_{\bm{\theta}_i} \ell_i (\bm{\theta}_i ; \mathcal{X}_n)$ is time consuming, so we do not recalculate $\bm{\hat\theta}^*_i = \argmax_{\bm{\theta}_i} \ell_i (\bm{\theta}_i ; \mathcal{X}^*_{n'})$
for bootstrap replicates.
Define the site-wise log-likelihood at site $t$ for tree T$i$ as
\begin{equation} \label{eq:sitewise-loglikelihood}
    \xi_{ti} = \log  p_i(\bm{x}_t ; \bm{\hat\theta}_i),\quad t=1,\ldots,n,\,i=1,\ldots, 105,
\end{equation}
so that the log-likelihood value for tree T$i$ is written as $\ell_i(\bm{\hat\theta}_i ; \mathcal{X}_n) = \sum_{t=1}^n \xi_{ti}$.
The bootstrap replicate of the log-likelihood value is approximated as
\begin{equation} \label{eq:rell}
    \ell_i(\bm{\hat\theta}^*_i ; \mathcal{X}^*_{n'}) \approx \ell_i(\bm{\hat\theta}_i ; \mathcal{X}^*_{n'}) = \sum_{t=1}^{n} w^*_t \xi_{ti},
\end{equation}
where $w^*_t$ is the number of times $\bm{x}_t$ appears in $\mathcal{X}^*_{n'}$.
The accuracy of this approximation as well as the higher-order term is given in eqs.~(4) and (5) of \cite{shimodaira2001multiple}.
Once $ \ell_i(\bm{\hat\theta}^*_i ; \mathcal{X}^*_{n'})$, $i=1,\ldots,105$, are computed by (\ref{eq:rell}), its ML tree is T$\hat i^*$ with $\hat i^* = \argmax_{i=1,\ldots,105} \ell_i(\bm{\hat\theta}^*_i ; \mathcal{X}^*_{n'})$.

The non-parametric bootstrap probability of tree T$i$ is obtained as follows. 
We generate $B$ bootstrap replicates $\bm{X}^{*b}_{n'}$, $b=1,\ldots, B$.  
In this paper, we used $B=10^5$.
For each $\bm{X}^{*b}_{n'}$, the ML tree T$\hat  i^{* b}$ is computed by the method described above. Then we count the frequency that T$i$ becomes the ML tree in the $B$ replicates.
The non-parametric bootstrap probability of tree T$i$ is computed by
\begin{equation} \label{eq:raw-bp-tree}
    \text{BP}(\text{T}i, n') = \#\{  \hat i^{* b} = i,\, b=1,\ldots, B\}/B.
\end{equation}
The non-parametric bootstrap probability of a edge is computed by summing $\text{BP}(\text{T}i, n') $ over the associated trees.

An example of the transformation $\bm{Y}^* = \bm{f}_n(\mathcal{X}^*_{n'})$ mentioned in Section~\ref{sec:multiscale_bootstrap} is
\begin{equation} \label{eq:fn-tree}
   \bm{Y}^* = \bm{V}_n^{-1/2} \bm{L}^*_{n'},
\end{equation}
where $\bm{L}^*_{n'}=(1/n')(\ell^*_1,\ldots,\ell^*_{105})^T$ with $\ell^*_i =  \ell_i(\bm{\hat\theta}^*_i ; \mathcal{X}^*_{n'}) $
and $\bm{V}_n$ is the variance matrix of $\bm{L}^*_n$. According to the approximation (\ref{eq:rell}) and the central limit theorem, (\ref{eq:y-msboot}) holds well for sufficiently large $n$ and $n'$ with $m=104$ and $\sigma^2=n/n'$.
It also follows from the above argument that $ \text{var}(\ell^*_i - \ell^*_j )  \approx  (n'/n) \|\bm{\xi}_i - \bm{\xi}_j\|^2$, and thus the variance of log-likelihood difference is
\begin{equation} \label{eq:var-lik}
   \text{var}\Bigl(\ell_i(\bm{\hat\theta}_i ; \mathcal{X}_{n}) - \ell_j(\bm{\hat\theta}_j ; \mathcal{X}_{n})   \Bigr) \approx \|\bm{\xi}_i - \bm{\xi}_j\|^2,
\end{equation}
which gives another insight into the visualization of Section~\ref{sec:model_map}, where the variance can be interpreted as the divergence between the two models; see eq.~(\ref{eq:var-divergence}).
This approximation holds well when the two predictive distributions $p_i(\bm{x};\bm{\hat\theta}_i)$, $p_j(\bm{x};\bm{\hat\theta}_j)$ are not very close to each other. 
When they are close to each other, however,
the higher-order term ignored in (\ref{eq:var-lik}) becomes dominant, and there is a difficulty for deriving the limiting distribution of the log-likelihood difference in the model selection test
\citep{shimodaira1997assessing,schennach2017simple}.

\subsection{Visualization of Probability Models} \label{sec:model_map}

\begin{figure}[h!]
\begin{center}
\includegraphics[width=15cm]{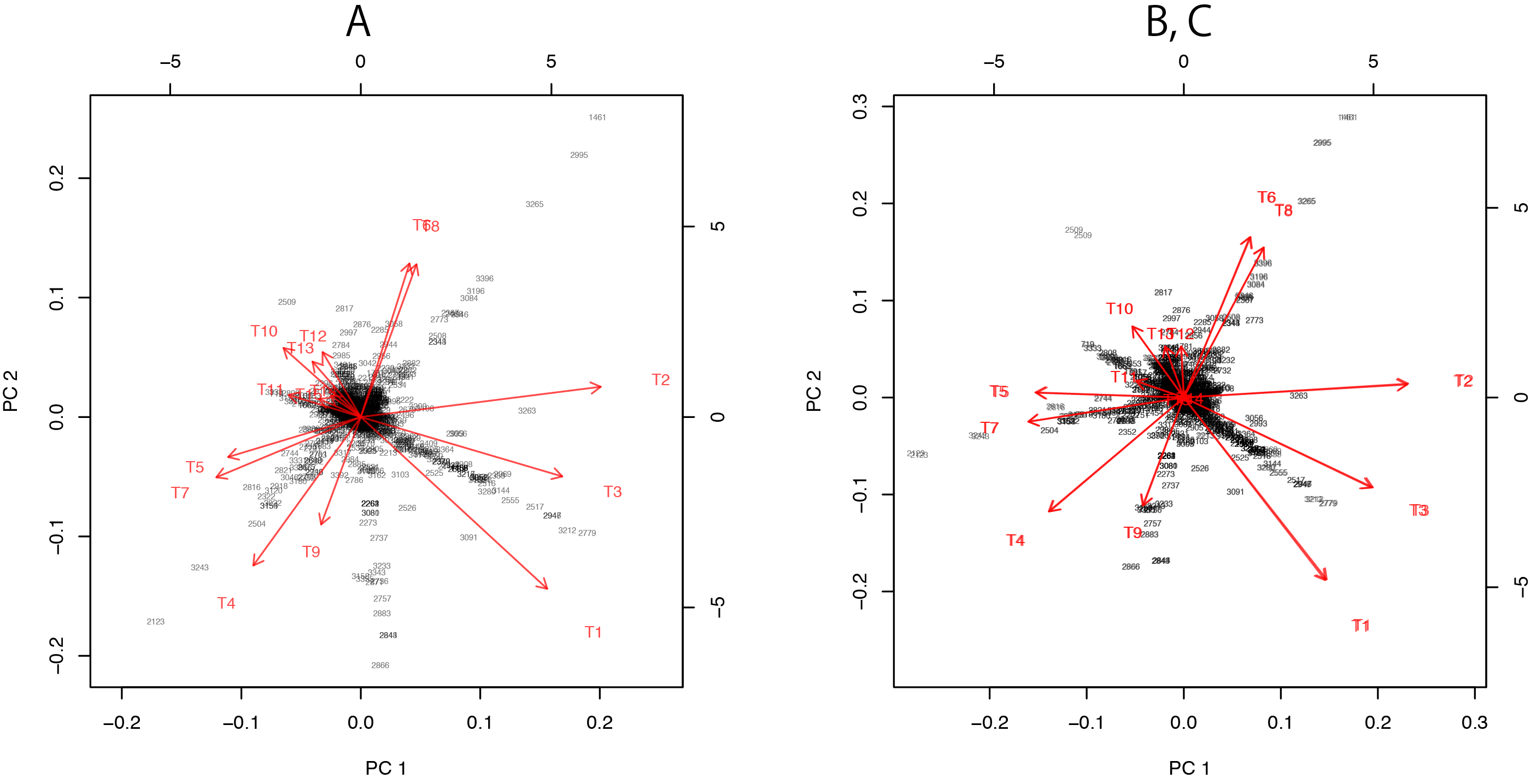}
\end{center}
\caption{Three versions the visualization of probability distributions for the best 15 trees drawn using different sets of models.
(A)~Only the 15 bifurcating trees. 
(B) 15 bifurcating trees + 10 partially resolved trees + 1 star topology.
This is the same plot as figure~\ref{fig:mam-models}B.
(C)~15 bifurcating trees + 1 star topology.
Note that B and C are superimposed, since their plots are almost indistinguishable.}\label{fig:mam-modelmaps}
\end{figure}

For representing the probability distribution of tree T$i$,
we define $\bm{\xi}_i := (\xi_{1i},\ldots, \xi_{ni})^T \in \mathbb{R}^n$ from (\ref{eq:sitewise-loglikelihood}) for $i=1,\ldots, 15$.
The idea behind the visualization of figure~\ref{fig:mam-models} is that
locations of $\bm{\xi}_i$ in $\mathbb{R}^n$ will represent locations of $p_i(\bm{x} ; \bm{\hat\theta}_i)$ in the space of probability distributions.  Let $D_\text{KL}(p_i \| p_j)$ be the Kullback-Leibler divergence between the two distributions.
For sufficiently small $(1/n) \|  \bm{\xi}_i - \bm{\xi}_j \|^2 $, the squared distance in $\mathbb{R}^n$ approximates $n$ times Jeffreys divergence
\begin{equation} \label{eq:var-divergence}
    \|  \bm{\xi}_i - \bm{\xi}_j \|^2 \approx n\times\Bigl( D_\text{KL}(p_i(\bm{x} ; \bm{\hat\theta}_i) \| p_j(\bm{x} ; \bm{\hat\theta}_j))
    + D_\text{KL}(p_j(\bm{x} ; \bm{\hat\theta}_j) \| p_i(\bm{x} ; \bm{\hat\theta}_i) \Bigr)
\end{equation}
for non-nested models~\citep[Section~6]{shimodaira2001multiple}.
When a model $p_0$ is nested in $p_i$, it becomes
$\|  \bm{\xi}_i - \bm{\xi}_0 \|^2 \approx
2n\times  D_\text{KL}(p_i(\bm{x} ; \bm{\hat\theta}_i) \| p_0(\bm{x} ; \bm{\hat\theta}_0) ) \approx
2\times (\ell_i(\bm{\hat\theta}_i ; \mathcal{X}_{n}) - \ell_0(\bm{\hat\theta}_0 ; \mathcal{X}_{n})) $.
We explain three different visualizations of figure~\ref{fig:mam-modelmaps}.
There are only minor differences between the plots, and the visualization is not sensitive to the details.

For dimensionality reduction, we have to specify the origin $\bm{c}\in\mathbb{R}^n$ and consider vectors $\bm{a}_i := \bm{\xi}_i - \bm{c}$.
A naive choice would be the average $\bm{c} = \sum_{i=1}^{15} \bm{\xi}_i/15$.
By applying PCA without centering and scaling (e.g., \texttt{prcomp} with option \texttt{center=FALSE, scale=FALSE} in R) to the matrix $(\bm{a}_1,\ldots,\bm{a}_{15})$, we obtain the visualization of $\bm{\xi}_i$ as the axes (red arrows) of biplot in figure~\ref{fig:mam-modelmaps}A.

For computing the ``data point'' $X$ in figure~\ref{fig:mam-models}, we need more models.
Let tree T106 be the star topology with no internal branch (completely unresolved tree), and T107$,\ldots,$ T131 be partially resolved tree topologies with only one internal branch corresponding to E1$,\ldots,$ E25, whereas T1$,\ldots,$ T105 are fully resolved trees (bifurcating trees).
Then define $\bm{\eta}_i := \bm{\xi}_{106+i}$, $i=0,\ldots, 25$.
Now we take $\bm{c}=\bm{\eta}_0$ for computing $\bm{a}_i = \bm{\xi}_i - \bm{\eta}_0$ and $\bm{b}_i = \bm{\eta}_i - \bm{\eta}_0$.
There is hierarchy of models: $\bm{\eta}_0$ is the submodel nested in all the other models, and $\bm{\eta}_1, \bm{\eta}_2, \bm{\eta}_3$, for example, are submodels of $\bm{\xi}_1$ (T1 includes E1, E2, E3).
By combining these non-nested models, we can reconstruct a comprehensive model in which all the other models are nested as submodels \citep[eq.~(10) in Section~5]{shimodaira2001multiple}.
The idea is analogous to reconstructing the full model $y=\beta_1 x_1 + \cdots + \beta_{25} x_{25} +\epsilon$ of multiple regression from submodels $y=\beta_1 x_1 + \epsilon,\ldots, y=\beta_{25} x_{25} + \epsilon$.
Thus we call it as ``full model'' in this paper, and the ML estimate of the full model is indicated as the data point $X$; it is also said ``super model'' in \cite{shimodaira2005assessing}.
Let $\bm{B}=(\bm{b}_1,\ldots,\bm{b}_{25})\in \mathbb{R}^{n \times 25}$ and $\bm{d} = (\|\bm{b}_1\|^2,\ldots, \|\bm{b}_{25}\|^2)^T \in \mathbb{R}^{25}$, then the vector for the full model is computed approximately by
\begin{equation} \label{eq:full-model}
    \bm{a}_X = \bm{B} (\bm{B}^T \bm{B})^{-1} \bm{d}.
\end{equation}
For the visualization of the best 15 trees, we may use only $\bm{b}_1,\ldots,\bm{b}_{11}$, because they include E1 and two more edges from E2$,\ldots,$E11.
In figures~\ref{fig:mam-models} and \ref{fig:mam-modelmaps}B, 
we actually modified the above computation slightly so that the star topology T106 is replaced by T107, the partially resolved tree corresponding to E1 (T107 is also said star topology by treating clade (23) as a leaf of the tree),
and the 10 partially resolved trees for E2$,\ldots,$ E11 are replaced by those for (E1,E2)$,\ldots,$ (E1,E11), respectively;
the origin becomes the maximal model nested in all the 15 trees, and $X$ becomes the minimal full model containing all the 15 trees.
Just before applying PCA in figure~\ref{fig:mam-modelmaps}B,
$\bm{a}_1,\ldots,\bm{a}_{15}$ are projected to the space orthogonal to $\bm{a}_X$, so that
the plot becomes the ``top-view'' of figure~\ref{fig:mam-models}A with $\bm{a}_X$ being at the origin.

In figure~\ref{fig:mam-modelmaps}C, we attempted a even simpler computation without using ML estimates for partially resolved trees. We used $\bm{B}=(\bm{a}_1,\ldots,\bm{a}_{15})$ and $\bm{d} = (\|\bm{a}_1\|^2,\ldots, \|\bm{a}_{15}\|^2)^T$, and taking the largest 10 singular values for computing the inverse in (\ref{eq:full-model}). The orthogonal projection to $\bm{a}_X$ is applied before PCA.

\subsection{Asymptotic Theory of Smooth Surfaces} \label{sec:smooth_surface}

For expressing the shape of the region $\mathcal{R}\subset\mathbb{R}^{m+1}$, we use a local coordinate system $(\bm{u},v)\in\mathbb{R}^{m+1}$ with $\bm{u}\in\mathbb{R}^m, v\in\mathbb{R}$. In a neighborhood of $\bm{y}$, the region is expressed as
\begin{equation} \label{eq:region-uv}
    \mathcal{R} = \{(\bm{u},v)  \mid  v \le - h(\bm{u}),\, \bm{u}\in\mathbb{R}^m  \},
\end{equation}
where $h$ is a smooth function; see \cite{shimodaira2008testing} for the theory of non-smooth surfaces. The boundary surface $\partial \mathcal{R}$ is expressed as $v=-h(\bm{u})$, $\bm{u}\in\mathbb{R}^m$.
We can choose the coordinates so that $\bm{y}=(\bm{0}, \beta_0)$ (i.e., $\bm{u}=(0,\ldots,0)$ and $v=\beta_0$), and $h(\bm{0})=0$, $\partial h/\partial u_i|_{\bm{0}} = 0$, $i=1,\ldots,m$.
The projection now becomes the origin $\bm{\hat\mu} = (\bm{0},0)$, and the signed distance is $\beta_0$. The mean curvature of surface $\partial \mathcal{R}$ at $\bm{\hat\mu}$ is now defined as
\begin{equation} \label{eq:beta1}
    \beta_1 = \frac{1}{2}\sum_{i=1}^m \frac{\partial^2 h(\bm{u})}{\partial u_i \partial u_i}\biggr|_{\bm{0}},
\end{equation}
which is interpreted as the trace of the hessian matrix of $h$.  When $\mathcal{R}$ is convex at least locally in the neighborhood, all the eigenvalues of the hessian are non-negative, leading to $\beta_1\ge0$, whereas concave $\mathcal{R}$ leads to $\beta_1\le0$.
In particular, $\beta_1=0$ when $\partial \mathcal{R}$ is flat (i.e., $h(\bm{u})\equiv 0$).

Since the transformation $\bm{y} = \bm{f}_n(\mathcal{X}_n)$ depends on $n$, the shape of the region $\mathcal{R}$ actually depends on $n$, although the dependency is implicit in the notation. As $n$ goes larger, the standard deviation of estimates, in general, reduces at the rate $n^{-1/2}$. For keeping the variance constant in (\ref{eq:y-model}), we actually magnifying the space by the factor $n^{1/2}$, meaning that the boundary surface $\partial \mathcal{R}$ approaches flat as $n\to\infty$. More specifically, the magnitude of  mean curvature is of order $\beta_1 = O_p(n^{-1/2})$. The magnitude of $\partial^3 h/\partial u_i \partial u_j \partial u_k$ and higher order derivatives is $O_p(n^{-1})$, and we ignore these terms in our asymptotic theory.
For keeping $\bm{\mu}=O(1)$ in  (\ref{eq:y-model}), we also consider the setting of ``local alternatives'', meaning that the parameter values approach a origin on the boundary at the rate $n^{-1/2}$.

\subsection{Bridging the Problem of Regions to the Z-Test} \label{sec:si_simple_case}

Here we explain the problem of regions in terms of the $z$-test by bridging the multivariate problem of Section~\ref{sec:theory} to the 1-dimensional case of Section~\ref{sec:intro}.

Ideal $p$-values are uniformly distributed over $p \in (0,1)$ when the null hypothesis holds. 
In fact, $\text{AU}(\mathcal{R}|\bm{Y}) \sim U(0,1)$ for $\bm{\mu}\in \partial \mathcal{R}$ as indicated in (\ref{eq:au-typeierror}).
The statistic  $\text{AU}(\mathcal{R}|\bm{Y})$  may be called \emph{pivotal} in the sense that the distribution does not change when $\bm{\mu}\in\partial\mathcal{R}$ moves on the surface.
Here we ignore the error of $O_p(n^{-1})$, and consider only the second order asymptotic accuracy.
From (\ref{eq:au-beta}), we can write $\text{AU}(\mathcal{R}|\bm{Y}) \simeq  \bar\Phi(\beta_0(\bm{Y}) - \beta_1(\bm{Y}) )$,
where the notation such as $\beta_0(\bm{Y})$ and $\beta_1(\bm{Y})$ indicates the dependency on $\bm{Y}$.
Since $\beta_1(\bm{Y})  \simeq \beta_1(\bm{y}) = \beta_1$, we treat $\beta_1(\bm{Y})$ as a constant.
Now we get the normal pivotal quantity~\citep{Efron:1985:BCI} as $\bar\Phi^{-1}( \text{AU}(\mathcal{R}|\bm{Y}) ) = \beta_0(\bm{Y}) - \beta_1 \sim N(0,1)$ for $\bm{\mu}\in \partial \mathcal{R}$.
More generally, it becomes
\begin{equation} \label{eq:pivot}
    \beta_0(\bm{Y}) - \beta_1 \sim N(\beta_0(\bm{\mu}),1),\quad \bm{\mu}\in\mathbb{R}^{m+1}.
\end{equation}

Let us look at the $z$-test in Section~\ref{sec:intro}, and consider substitutions:
\begin{equation} \label{eq:bridge}
    Z = \beta_0(\bm{Y}) - \beta_1,\quad \theta = \beta_0(\bm{\mu}),\quad c = -\beta_1.
\end{equation}
The 1-dimensional model (\ref{eq:z-normal}) is now equivalent to (\ref{eq:pivot}).
The null hypothesis is also equivalent: $\theta\le 0 \Leftrightarrow \beta_0(\bm{\mu}) \le 0 \Leftrightarrow \bm{\mu}\in\mathcal{R}$.
We can easily verify that AU corresponds to $p(z)$, because  $p(z) = \bar\Phi(z) = \bar\Phi(\beta_0(\bm{y}) - \beta_1) \simeq \text{AU}(\mathcal{R}|\bm{y}) $, which is expected from the way we obtained (\ref{eq:pivot}) above.
Furthermore, we can derive SI from $p(z,c)$.
First verify that the selection event is equivalent: $Z > c \Leftrightarrow \beta_0(\bm{Y}) - \beta_1 > -\beta_1 \Leftrightarrow  \beta_0(\bm{Y})  > 0 \Leftrightarrow \bm{Y} \in \mathcal{R}^C$.
Finally, we obtain SI as
$p(z,c) = p(z)/\bar\Phi(c) \simeq   \bar\Phi(\beta_0(\bm{y}) - \beta_1) /\bar\Phi(-\beta_1) \simeq  \text{SI}(\mathcal{R}|\bm{y})   $.

\subsection{Model Fitting in Multiscale Bootstrap} \label{sec:msboot-detail}

\begin{figure}[h!]
\begin{center}
\includegraphics[width=15cm]{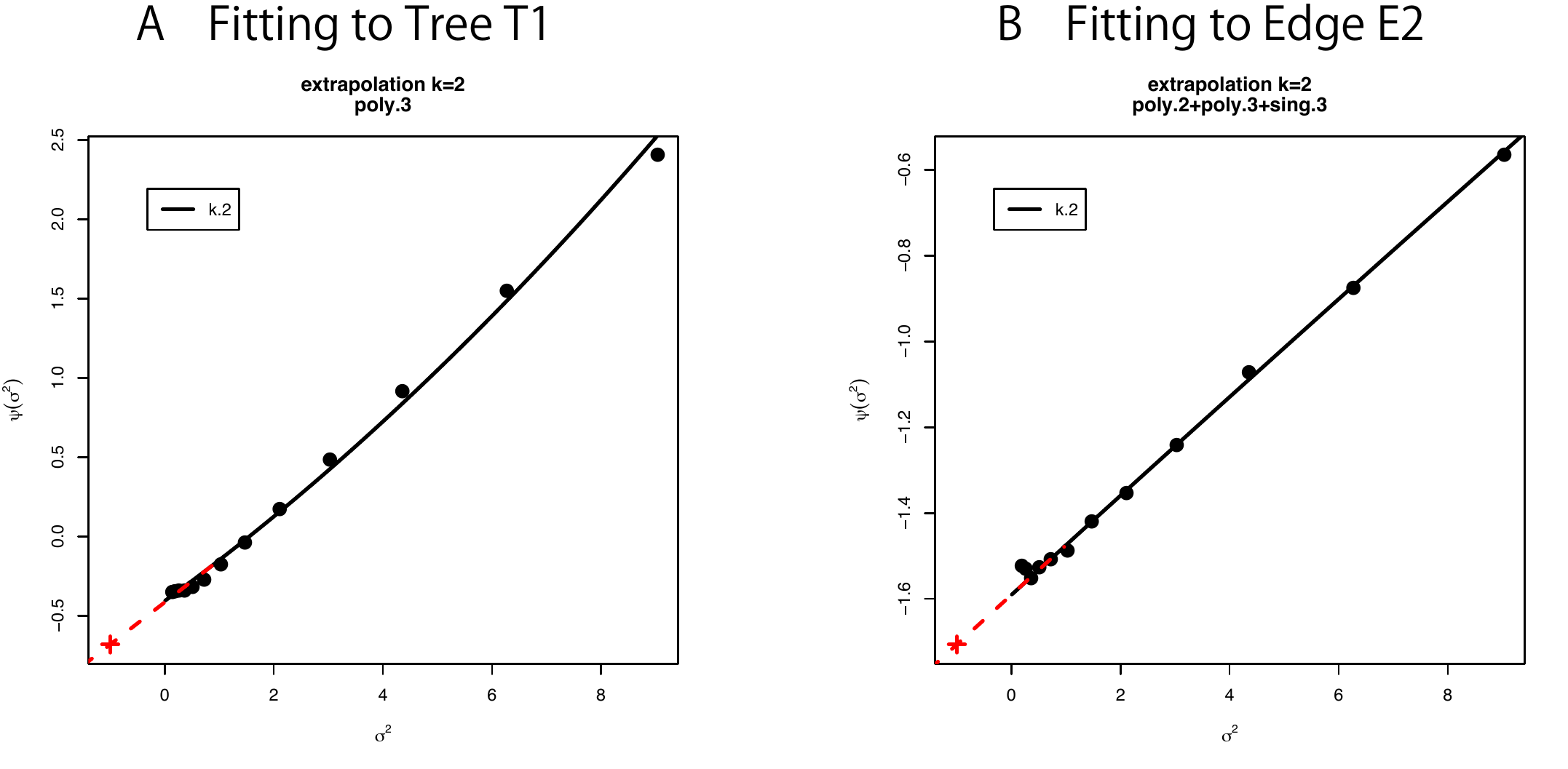}
\end{center}
\caption{Multiscale bootstrap for (A)~tree T1 and (B)~edge E2.
$\psi_{\sigma^2}(\mathcal{R}|\bm{y})$ is computed by the non-parametric bootstrap probabilities for several $\sigma^2=n/n'$ values,
then $\beta_0$ and $\beta_1$ are estimated as the intercept and the slope, respectively. See Section~\ref{sec:msboot-detail} for details.
}\label{fig:mam-fitting}
\end{figure}

We have used thirteen $\sigma^2$ values from 1/9 to 9 (equally spaced in log-scale). This range is relatively large, and we observe a slight deviation from the linear model $\beta_0 + \beta_1 \sigma^2$ in figure~\ref{fig:mam-fitting}.
Therefore we fit other models to the observed values of $\psi_{\sigma^2}$ as implemented in \emph{scaleboot} package~\citep{shimodaira2008testing}.
For example, poly.$k$ model is $\sum_{i=0}^{k-1} \beta_i \sigma^{2i}$, and sing.3 model is $\beta_0 + \beta_1 \sigma^2 (1 + \beta_2(\sigma-1))^{-1}$. In figure~\ref{fig:mam-fitting}A, poly.3 is the best model according to AIC \citep{akaike1974new}.
In figure~\ref{fig:mam-fitting}B, poly.2, poly.3, and sing.3 are combined by model averaging with Akaike weights.
Then $\beta_0$ and $\beta_1$ are estimated from the tangent line to the fitted curve of $\psi_{\sigma^2}$ at $\sigma^2=1$.
In figure~\ref{fig:mam-fitting}, the tangent line is drawn as red line for extrapolating $\psi_{\sigma^2}$ to $\sigma^2=-1$.
\cite{shimodaira2008testing,terada2017selective} considered the Taylor expansion of $\psi_{\sigma^2}$ at $\sigma^2=1$ as a generalization of the tangent line for improving the accuracy of AU and SI.

In the implementation of \emph{CONSEL}~\citep{shimodaira2001consel} and $\emph{pvclust}$~\citep{suzuki2006pvclust}, we use a narrower range of $\sigma^2$ values (ten $\sigma^{-2}$ values: 0.5, 0.6, $\ldots,$ 1.4). Only the linear model $\beta_0 + \beta_1 \sigma^2$ is fitted there. The estimated $\beta_0$ and $\beta_1$ should be very close to those estimated from the tangent line described above.
An advantage of using wider range of $\sigma^2$ in \emph{scaleboot} is that the standard error of $\beta_0$ and $\beta_1$  will become smaller.

\subsection{General Formula of Selective Inference} \label{sec:si_general}

Let $\mathcal{H}, \mathcal{S} \subset \mathbb{R}^{m+1}$ be regions for the null hypothesis and the selection event, respectively.
We would like to test the null hypothesis $H_0: \bm{\mu}\in\mathcal{H}$ against the alternative  $H_1: \bm{\mu}\in\mathcal{H}^C$ conditioned on the selection event $\bm{y}\in\mathcal{S}$. 
We have considered the outside mode $\mathcal{H}=\mathcal{R}, \mathcal{S}=\mathcal{R}^C$ in (\ref{eq:si-beta}) and the inside mode $\mathcal{H}=\mathcal{R}^C, \mathcal{S}=\mathcal{R}$ in (\ref{eq:si-def-beta-c}).
For a general case of $\mathcal{H}, \mathcal{S}$, \cite{terada2017selective} gave a formula of approximately unbiased $p$-value of selective inference as
\begin{equation} \label{eq:si_general}
    \text{SI}( \mathcal{H} | \mathcal{S}, \bm{y} ) 
    = \frac{\bar\Phi(\beta_0^\mathcal{H} - \beta_1^\mathcal{H})}{\bar\Phi(\beta_0^\mathcal{S} + \beta_0^\mathcal{H} - \beta_1^\mathcal{H})},
\end{equation}
where geometric quantities $\beta_0, \beta_1$ are defined for the regions $\mathcal{H}, \mathcal{S}$.
We assumed that $\mathcal{H}$ and $\mathcal{S}^C$ are expressed as (\ref{eq:region-uv}), and two surfaces $\partial \mathcal{H}, \partial \mathcal{S}$ are nearly parallel to each other with tangent planes differing only $O_p(n^{-1/2})$.
The last assumption always holds for (\ref{eq:si-beta}), because $\partial \mathcal{H} = \partial \mathcal{R}$ and  $\partial \mathcal{S} = \partial \mathcal{R}^C$ are identical and of course parallel to each other.

Here we explain why we have considered the special case of $\mathcal{S}=\mathcal{H}^C$ for phylogenetic inference.
First, we suppose that the selection event satisfies $\mathcal{S}\subset\mathcal{H}^C$, because a reasonable test would not reject $H_0$ unless $\bm{y}\in\mathcal{H}^C$.
Note that $\bm{y}\in \mathcal{S}\subset\mathcal{H}^C$ implies $0 \le - \beta_0^\mathcal{S}  \le  \beta_0^\mathcal{H}$.
Therefore, $\beta_0^\mathcal{H} + \beta_0^\mathcal{S} \ge 0$ leads to
\begin{equation}
    \text{SI}( \mathcal{H} | \mathcal{S}, \bm{y} )  \ge \text{SI}( \mathcal{H} |  \bm{y} ),
\end{equation}
where $\text{SI}( \mathcal{H} |  \bm{y} ) := \text{SI}( \mathcal{H} | \mathcal{H}^C, \bm{y} )$  is obtained from (\ref{eq:si_general}) by letting $\beta_0^\mathcal{H}  + \beta_0^\mathcal{S} = 0$ for $\mathcal{S}=\mathcal{H}^C$.
The $p$-value $\text{SI}( \mathcal{H} | \mathcal{S}, \bm{y} )$ becomes smaller as $\mathcal{S}$ grows, and $\mathcal{S}=\mathcal{H}^C$ gives the smallest $p$-value, leading to the most powerful selective test.
Therefore the choice $\mathcal{S}=\mathcal{H}^C$ is preferable to any other choice of selection event satisfying $\mathcal{S}\subset\mathcal{H}^C$. 
This kind of property is mentioned in \cite{fithian2014optimal} as the monotonicity of selective error in the context of ``data curving''.

Let us see how these two $p$-values differ for the case of E2 by specifying
$\mathcal{H}=\mathcal{R}^C_\text{E2}$ and $\mathcal{S}=\mathcal{R}_\text{T1}$.
In this case, the two surfaces $\partial \mathcal{H}, \partial \mathcal{S}$ may not be very parallel to each other, thus violating the assumption of $\text{SI}( \mathcal{H} | \mathcal{S}, \bm{y} )$, so we only intend to show the potential difference between the two $p$-values.
The geometric quantities are
$\beta_0^\mathcal{H} = -\beta_0^\text{E2} = 1.59$,
$\beta_1^\mathcal{H} = -\beta_1^\text{E2} = -0.12$,
$\beta_0^\mathcal{S} = \beta_0^\text{T1} = -0.41$;
the $p$-values are calculated using more decimal places than shown.
SI of E2 conditioned on selecting T1 is
$$
\text{SI}( \mathcal{H} | \mathcal{S}, \bm{y} ) = \frac{\bar\Phi(1.59+0.12)}{\bar\Phi(-0.41+1.59+0.21)} = 0.448,
$$
and it is very different from SI of E2 conditioned on selecting E2
$$
\text{SI}( \mathcal{H} |\bm{y} ) = \frac{\bar\Phi(1.59+0.12)}{\bar\Phi(0.12)} = 0.097,
$$
where $\text{SI}'(\mathcal{R}_\text{E2}^C | \bm{y}) = 1 - \text{SI}(\mathcal{R}_\text{E2}^C | \bm{y})=0.903$ is shown in table~\ref{tab:mam-edges}.
As you see, $\text{SI}( \mathcal{H} |\bm{y} )$ is easier to reject $H_0$ than $\text{SI}( \mathcal{H} | \mathcal{S}, \bm{y} )$.

\subsection{Number of regions for phylogenetic inference} \label{sec:number_regions}

The regions $\mathcal{R}_i$, $i=1,\ldots, K_\text{all}$ correspond to trees or edges.
In inside and outside modes, the number of total regions is
$K_\text{all}=105$ for trees and $K_\text{all}=25$ for edges when the number of taxa is $N=6$. 
For general $N\ge3$, they grow rapidly as $K_\text{all} = (2N-5)!/(2^{N-3}(N-3)!)$ for trees and $K_\text{all} = 2^{N-1} - (N+1)$ for edges. 
Next consider the number of selected regions $K_\text{select}$.
In inside mode, regions with $\bm{y}\in\mathcal{R}_i$ are selected, and the number is counted as  $K_\text{select}=1$ for trees and $K_\text{select}=N-3=3$ for edges.
In outside mode, regions with $\bm{y}\not\in\mathcal{R}_i$ are selected, and thus the number is $K_\text{all}$ minus that for inside mode; 
$K_\text{select}=K_\text{all}-1=104$ for trees and $K_\text{select}=K_\text{all}-(N-3)=22$ for edges.
Finally, consider the number of true null hypotheses, denoted as $K_\text{true}$.
The null hypothesis holds true when $\bm{\mu}\not\in\mathcal{R}_i$ in inside mode and  $\bm{\mu}\in\mathcal{R}_i$ in outside mode, and thus $K_\text{true}$ is the same as the number of regions with  $\bm{y}\not\in\mathcal{R}_i$ in inside mode and  $\bm{y}\in\mathcal{R}_i$ in outside mode (These numbers do not depend on the value of $\bm{y}$ by ignoring the case of $\bm{y}\in\partial\mathcal{R}_i$).
Therefore, $K_\text{true} = K_\text{all} - K_\text{select}$ for both cases.

\subsection{Selective Inference of Lasso Regression} \label{sec:si_lasso}

Selective inference is considered for the variable selection of regression analysis.
Here, we deal with prostate cancer data \citep{StameyEtAl89}
in which we predict the level of prostate-specific antigen (PSA) 
from clinical measures.
The dataset is available in the R package \emph{ElemStatLearn} \citep{ElemStatLearn2015}.
We consider a linear model to the log of PSA (\texttt{lpsa}), with $8$ predictors such as the log prostate weight (\texttt{lweight}), \texttt{age}, and so on.
All the variables are standardized to have zero mean and unit variance.

The goal is to provide the valid selective inference for 
the partial regression coefficients of the selected variables by lasso~\citep{tibshirani1996regression}.
Let $n$ and $p$ be the number of observations and the number of predictors.
$\bm{\hat{M}}$ is the set of selected variables,
and $\bm{\hat{s}}$ represents the signs of the selected regression coefficients.
We suppose that regression responses are distributed as $\bm{Y} \sim N(\bm{\mu},\tau^2 \bm{I}_n)$ where
$\bm{\mu} \in \mathbb{R}^n$ and $\tau>0$.
Let $e_i$ be the $i$th residual. Resampling the scaled residuals $\sigma e_i\;(i=1,\dots,n)$ with several values of scale $\sigma^2$, 
we can apply the multiscale bootstrap method described in Section~\ref{sec:selective_inference}
for the selective inference in the regression problem.
Here, we note that the target of the inference is the true partial regression coefficients:
\[
\bm{\beta} = (\bm{X}^T\bm{X})^{-1}\bm{X}^T\bm{\mu},
\]
where $\bm{X}\in \mathbb{R}^{n\times p}$ is the design matrix.
We compute four types of intervals with confidence level $1-\alpha=0.95$ for selected variable $j$.
$[L_j^\text{ordinary}, U_j^\text{ordinary}]$
is the non-selective confidence interval obtained via $t$-distribution.
$[L_j^\text{model}, U_j^\text{model}]$
is the selective confidence interval under the selected model proposed by \cite{LeeEtAl16} and \cite{RyanTibshiraniEtAl16},
which is computed by \texttt{fixedLassoInf} with \texttt{type="full"} in R package \emph{selectiveInference} \citep{selectiveInference2017}.
By extending the method of $[L_j^\text{model}, U_j^\text{model}]$, we also computed
$[L_j^\text{variable}, U_j^\text{variable}]$, which
is the selective confidence interval under the selection event that variable $j$ is selected.
These three confidence intervals are exact, in the sense that
\begin{align*}
&P\left( \beta_j \in[L_j^\text{ordinary}, U_j^\text{ordinary}]  \right)=1-\alpha, \quad
P\left( \beta_j \in [L_j^\text{model}, U_j^\text{model}] \mid \bm{\hat{M}}, \bm{\hat{s}} \right)=1-\alpha,\\
&P\left( \beta_j \in[L_j^\text{variable}, U_j^\text{variable}] \mid j\in \bm{\hat{M}}, \hat{s}_j \right)=1-\alpha.
\end{align*}
Note that the selection event of variable $j$, i.e., $\{j\in \bm{\hat{M}}, \hat{s}_j \}$ can be represented as a union of polyhedra on $\mathbb{R}^n$, and thus, according to the polyhedral lemma \citep{LeeEtAl16, RyanTibshiraniEtAl16}, we can compute a valid confidence interval $[L_j^\text{variable}, U_j^\text{variable}]$.
However, this computation is prohibitive for $p > 10$,
because all the possible combinations of models with variable $j$ are considered.
Therefore, we compute its approximation
$[\hat L_j^\text{variable}, \hat U_j^\text{variable}]$
by the multiscale bootstrap method
of Section~\ref{sec:selective_inference}
with much faster computation even for larger $p$.

We set $\lambda = 10$ as the penalty parameter of lasso, and 
the following model and signs were selected:
\[
\bm{\hat M} = \{\texttt{lcavol},\texttt{lweight},\texttt{lbph},\texttt{svi},\texttt{pgg45}\},\quad \bm{\hat s} = (+,+,+,+,+).
\]
The confidence intervals are shown in figure~\ref{fig:lasso}.
For adjusting the selection bias, the three confidence intervals of selective inference are longer than the ordinary confidence interval.
Comparing $[L_j^\text{model}, U_j^\text{model}]$ and $[ L_j^\text{variable}, U_j^\text{variable}]$,
the latter is shorter, and would be preferable.
This is because the selection event of the latter is less restrictive as $\{\bm{\hat{M}}, \bm{\hat{s}}\}\subseteq \{j\in \bm{\hat{M}}, \hat{s}_j \}$; see Section~\ref{sec:si_general} for the reason why larger selection event is better.
Finally, we verify that  $[ \hat L_j^\text{variable}, \hat U_j^\text{variable}]$ approximates  $[ L_j^\text{variable}, U_j^\text{variable}]$ very well.

\section*{Conflict of Interest Statement}

The authors declare that the research was conducted in the absence of any commercial or financial relationships that could be construed as a potential conflict of interest.

\section*{Author Contributions}

HS and YT developed the theory of selective inference. HS programmed the multiscale bootstrap software and conducted the phylogenetic analysis. YT conducted the lasso analysis. HS wrote the manuscript. All authors have approved the final version of the manuscript.


\section*{Funding}

This research was supported in part by JSPS KAKENHI Grant (16H02789 to HS, 16K16024 to YT).


\section*{Acknowledgments}
The authors appreciate the feedback from the audience of seminar talk of HS at Department of Statistics, Stanford University.
The authors are grateful to Masami Hasegawa for his insightful comments on phylogenetic analysis of mammal species.


\section*{Data Availability Statement}
The datasets analyzed for this study can be found in the software package \emph{scaleboot}
\citep{shimo2019scaleboot}.



\bibliographystyle{imsart-nameyear}
\bibliography{stat2019}





\clearpage

\section*{Figure captions}

\clearpage

\section*{Tables}

\newcommand{\sig}[1]{\textbf{#1}}
\newcommand{\siga}[1]{\textbf{\underline{#1}}}

\begin{table}[h!]
\begin{center}
\caption{Three types of $p$-values (BP, AU, SI) and geometric quantities ($\beta_0, \beta_1$) for the best 20 trees.
Standard errors are shown in parentheses.
Boldface indicates significance ($p<0.05$) for the null hypothesis that the tree is true (outside mode).
For the rest of trees (T21$,\ldots,$ T105), $p$-values are very small ($p < 0.001$).
}\label{tab:mam-trees}
\vskip1em
\renewcommand{\arraystretch}{1.1}
\tiny
\begin{tabular}{lcccD{.}{.}{0}lD{.}{.}{0}lll}
\hline
tree & BP & AU & SI & \multicolumn{2}{c}{$\beta_0$} & \multicolumn{2}{c}{$\beta_1$} & topology & edges \\
\hline
 T1$^\dagger$ & 0.559 (0.001) & 0.752 (0.001) & 0.372 (0.001) & -0.41 & (0.00) & 0.27 & (0.00) & (((1(23))4)56) & E1,E2,E3  \\ 
 
 T2 & 0.304 (0.000) & 0.467 (0.001) & 0.798 (0.001) &  0.30 & (0.00) & 0.22 & (0.00) & ((1((23)4))56) & E1,E2,E4  \\ 
 T3 & \sig{0.038} (0.000) & 0.126 (0.002) & 0.202 (0.003) &  1.46 & (0.01) & 0.32 & (0.00) & (((14)(23))56) & E1,E2,E5  \\ 
 T4 & \sig{0.014} (0.000) & 0.081 (0.002) & 0.124 (0.003) &  1.79 & (0.01) & 0.40 & (0.01) & ((1(23))(45)6) & E1,E3,E6  \\ 
 T5 & \sig{0.032} (0.000) & 0.127 (0.002) & 0.199 (0.003) &  1.50 & (0.01) & 0.36 & (0.00) & (1((23)(45))6) & E1,E6,E7  \\ 
 T6 & \sig{0.005} (0.000) & \sig{0.032} (0.002) & 0.050 (0.002) &  2.21 & (0.02) & 0.35 & (0.01) & (1(((23)4)5)6) & E1,E4,E7  \\ 
 T7$^\ddagger$ & \sig{0.015} (0.000) & 0.100 (0.003) & 0.150 (0.003) &  1.72 & (0.01) & 0.44 & (0.01) & ((1(45))(23)6) & E1,E6,E8  \\ 
 T8 & \sig{0.001} (0.000) & \sig{0.011} (0.001) & \sig{0.016} (0.002) &  2.74 & (0.03) & 0.43 & (0.02) & ((15)((23)4)6) & E1,E4,E9  \\ 
 T9 & \sig{0.000} (0.000) & \sig{0.001} (0.000) & \sig{0.001} (0.000) &  3.67 & (0.09) & 0.46 & (0.04) & (((1(23))5)46) & E1,E3,E10 \\ 
T10 & \sig{0.002} (0.000) & \sig{0.022} (0.002) & \sig{0.033} (0.002) &  2.43 & (0.02) & 0.42 & (0.01) & (((15)4)(23)6) & E1,E8,E9  \\ 
T11 & \sig{0.000} (0.000) & \sig{0.004} (0.001) & \sig{0.006} (0.002) &  3.14 & (0.07) & 0.51 & (0.03) & (((14)5)(23)6) & E1,E5,E8  \\ 
T12 & \sig{0.000} (0.000) & \sig{0.000} (0.000) & \sig{0.001} (0.000) &  3.78 & (0.09) & 0.41 & (0.04) & (((15)(23))46) & E1,E9,E10 \\ 
T13 & \sig{0.000} (0.000) & \sig{0.000} (0.000) & \sig{0.001} (0.001) &  3.96 & (0.19) & 0.54 & (0.09) & (1(((23)5)4)6) & E1,E7,E11 \\ 
T14 & \sig{0.000} (0.000) & \sig{0.000} (0.000) & \sig{0.000} (0.000) &  4.66 & (0.31) & 0.65 & (0.12) & ((14)((23)5)6) & E1,E5,E11 \\ 
T15 & \sig{0.000} (0.000) & \sig{0.000} (0.000) & \sig{0.000} (0.000) &  5.28 & (0.34) & 0.43 & (0.11) & ((1((23)5))46) & E1,E10,E11\\ 
T16 & \sig{0.000} (0.000) & \sig{0.000} (0.000) & \sig{0.001} (0.000) &  3.63 & (0.04) & 0.23 & (0.01) & ((((13)2)4)56) & E2,E3,E12 \\ 
T17 & \sig{0.000} (0.000) & \sig{0.000} (0.000) & \sig{0.000} (0.000) &  3.81 & (0.04) & 0.22 & (0.01) & ((((12)3)4)56) & E2,E3,E13 \\ 
T18 & \sig{0.000} (0.000) & \sig{0.000} (0.000) & \sig{0.000} (0.000) &  4.33 & (0.10) & 0.34 & (0.03) & (((13)2)(45)6) & E3,E6,E12 \\ 
T19 & \sig{0.000} (0.000) & \sig{0.000} (0.000) & \sig{0.000} (0.000) &  4.36 & (0.11) & 0.32 & (0.04) & (((12)3)(45)6) & E3,E6,E13 \\ 
T20 & \sig{0.000} (0.000) & \sig{0.000} (0.000) & \sig{0.000} (0.000) &  3.90 & (0.12) & 0.44 & (0.05) & (((1(45))2)36) & E6,E8,E14 \\ 
\hline
\end{tabular}
\end{center}
$^\dagger$ T1 is the ML tree, i.e., the tree selected by the ML method based on the dataset of \cite{shimodaira1999multiple}.
$^\ddagger$ T7 is presumably the true tree as suggested by later researches; see Section~\ref{sec:si_phylo}.

\end{table}

\begin{table}[h!]
\begin{center}
\caption{Three types of $p$-values (BP, AU, SI) and geometric quantities ($\beta_0, \beta_1$) for all the 25 edges of six taxa.
Standard errors are shown in parentheses.
Boldface without underline indicates significance ($p<0.05$) for the null hypothesis that the edge is true (outside mode).
Boldface with underline indicates significance ($p>0.95$) for the null hypothesis that the edge is \emph{not} true (inside mode).
}\label{tab:mam-edges}
\renewcommand{\arraystretch}{1.1}
\vskip1em
\tiny
\begin{tabular}{lcccD{.}{.}{0}lD{.}{.}{0}ll}
\hline
edge & BP & AU & SI & \multicolumn{2}{c}{$\beta_0$} & \multicolumn{2}{c}{$\beta_1$}  & clade \\
\hline
 E1$^{\dagger\ddagger}$ & \siga{1.000} (0.000) & \siga{1.000} (0.000) & \siga{1.000} (0.000) & -3.87 & (0.03) & 0.16 & (0.01) & \texttt{-++---} \\
 E2$^\dagger$  & 0.930 (0.000) & \siga{0.956} (0.001) & 0.903 (0.001) & -1.59 & (0.00) & 0.12 & (0.00) & \texttt{++++--} \\
 E3$^\dagger$  & 0.580 (0.001) & 0.719 (0.001) & 0.338 (0.001) & -0.39 & (0.00) & 0.19 & (0.00) & \texttt{+++---} \\
 
 E4 & 0.318 (0.000) & 0.435 (0.001) & 0.775 (0.001) &  0.32 & (0.00) & 0.16 & (0.00) & \texttt{-+++--} \\
 E5 & \sig{0.037} (0.000) & 0.124 (0.002) & 0.198 (0.002) &  1.47 & (0.01) & 0.32 & (0.00) & \texttt{+--+--} \\
 E6$^\ddagger$ & 0.060 (0.000) & 0.074 (0.001) & 0.141 (0.002) &  1.50 & (0.00) & 0.05 & (0.00) & \texttt{---++-} \\
 E7 & \sig{0.038} (0.000) & 0.091 (0.002) & 0.154 (0.002) &  1.56 & (0.01) & 0.22 & (0.00) & \texttt{-++++-} \\
 E8$^\ddagger$ & \sig{0.018} (0.000) & 0.068 (0.002) & 0.110 (0.003) &  1.80 & (0.01) & 0.31 & (0.01) & \texttt{+--++-} \\
 E9 & \sig{0.003} (0.000) & \sig{0.014} (0.001) & \sig{0.023} (0.002) &  2.48 & (0.02) & 0.27 & (0.02) & \texttt{+---+-} \\
E10 & \sig{0.000} (0.000) & \sig{0.000} (0.000) & \sig{0.001} (0.000) &  3.72 & (0.07) & 0.29 & (0.03) & \texttt{+++-+-} \\
E11 & \sig{0.000} (0.000) & \sig{0.000} (0.000) & \sig{0.000} (0.000) &  4.31 & (0.10) & 0.35 & (0.03) & \texttt{-++-+-} \\
E12 & \sig{0.000} (0.000) & \sig{0.000} (0.000) & \sig{0.000} (0.000) &  3.68 & (0.05) & 0.17 & (0.02) & \texttt{+-+---} \\
E13 & \sig{0.000} (0.000) & \sig{0.000} (0.000) & \sig{0.000} (0.000) &  3.90 & (0.04) & 0.15 & (0.02) & \texttt{++----} \\
E14 & \sig{0.000} (0.000) & \sig{0.000} (0.000) & \sig{0.000} (0.000) &  4.03 & (0.09) & 0.30 & (0.04) & \texttt{++-++-} \\
E15 & \sig{0.000} (0.000) & \sig{0.000} (0.000) & \sig{0.000} (0.000) &  4.03 & (0.13) & 0.38 & (0.06) & \texttt{+-+++-} \\
E16 & \sig{0.000} (0.000) & \sig{0.000} (0.000) & \sig{0.000} (0.000) &  4.44 & (0.05) & 0.12 & (0.01) & \texttt{-+-+--} \\
E17 & \sig{0.000} (0.000) & \sig{0.000} (0.000) & \sig{0.000} (0.000) &  4.70 & (0.07) & 0.19 & (0.02) & \texttt{++-+--} \\
E18 & \sig{0.000} (0.000) & \sig{0.000} (0.000) & \sig{0.000} (0.000) &  3.94 & (0.09) & 0.26 & (0.04) & \texttt{-+-++-} \\
E19 & \sig{0.000} (0.000) & \sig{0.000} (0.000) & \sig{0.000} (0.000) &  5.23 & (0.43) & 0.57 & (0.13) & \texttt{--++--} \\
E20 & \sig{0.000} (0.000) & \sig{0.000} (0.000) & \sig{0.000} (0.000) &  5.66 & (0.29) & 0.28 & (0.09) & \texttt{+-++--} \\
E21 & \sig{0.000} (0.000) & \sig{0.000} (0.000) & \sig{0.000} (0.000) &  6.38 & (0.33) & 0.24 & (0.08) & \texttt{--+++-} \\
E22 & \sig{0.000} (0.000) & \sig{0.000} (0.000) & \sig{0.000} (0.000) &  5.62 & (0.21) & 0.17 & (0.07) & \texttt{--+-+-} \\
E23 & \sig{0.000} (0.000) & \sig{0.000} (0.000) & \sig{0.000} (0.000) &  4.86 & (0.43) & 0.70 & (0.13) & \texttt{-+--+-} \\
E24 & \sig{0.000} (0.000) & \sig{0.000} (0.000) & \sig{0.000} (0.000) &  5.61 & (0.17) & 0.23 & (0.04) & \texttt{+-+-+-} \\
E25 & \sig{0.000} (0.000) & \sig{0.000} (0.000) & \sig{0.000} (0.000) &  6.32 & (0.71) & 0.52 & (0.20) & \texttt{++--+-} \\
\hline
\end{tabular}
\end{center}
$^\dagger$ Edges included in T1.
$^\ddagger$ Edges included in T7.

\end{table}

\begin{table}[h!]
\begin{center}
\caption{The number of regions for trees and edges. The number of taxa is $N=6$.}\label{tab:Knumbers}
\vskip1em
\renewcommand{\arraystretch}{1.3}
\begin{tabular}{lccccc}
\hline
  &\multicolumn{2}{c}{inside mode} & & \multicolumn{2}{c}{outside mode}\\
  \cline{2-3} \cline{5-6}
  & tree & edge & & tree & edge \\
\hline
 $K_\text{select}$ & 1 & 3 & & 104 & 22\\
 $K_\text{true}$ & 104 & 22 & & 1 & 3 \\
 $K_\text{all}$ & 105 & 25 & & 105 & 25\\
\hline 
\end{tabular}
\end{center}
\end{table}

\end{document}